\documentclass[fleqn,usenatbib]{mnras}

\usepackage[T1]{fontenc}
\usepackage{ae,aecompl}

\usepackage{graphicx}	
\usepackage{amsmath}	
\usepackage{amssymb}	
\usepackage{subfigure}

\newcommand{\gfil}{$g^{\prime}$}	
\newcommand{\rfil}{$r^{\prime}$}	
\newcommand\0{\phantom{0}}          

\title[Polarimetric and Radiative Transfer Modelling of HD~172555]{Polarimetric and Radiative Transfer Modelling of HD~172555}

\author[J. P. Marshall \textit{et al.}]{Jonathan P. Marshall$^{1,2}$\thanks{Contact e-mail: \href{mailto:jmarshall@asiaa.sinica.edu.tw}{jmarshall@asiaa.sinica.edu.tw}}, Daniel V. Cotton$^{2,3,4}$, Peter Scicluna$^{5, 1}$, \newauthor 
Jeremy Bailey$^{6}$, Lucyna Kedziora-Chudczer$^{2}$, and Kimberly Bott$^{7}$
\\
$^{1}$Academia Sinica, Institute of Astronomy and Astrophysics, 11F Astronomy-Mathematics Building, NTU/AS campus,\\ No. 1, Section 4, Roosevelt Rd., Taipei 10617, Taiwan\\
$^{2}$University of Southern Queensland, Centre for Astrophysics, Toowoomba, QLD 4350, Australia\\
$^{3}$Anglo Australian Telescope, Australian National University, 418 Observatory Road, Coonabarabran, NSW 2357, Australia.\\
$^{4}$Western Sydney University, Locked Bag 1797, Penrith-South DC, NSW 1797, Australia.\\
$^{5}$European Southern Observatory, Alonso de Cordova 3107, Santiago RM, Chile\\
$^{6}$School of Physics, University of New South Wales, Sydney, NSW 2052, Australia\\
$^{7}$Department of Earth and Planetary Science, University of California, Riverside, CA, 92521, USA}

\date{Accepted XXX. Received YYY; in original form ZZZ}

\pubyear{2020}

\begin{document}
\label{firstpage}
\pagerange{\pageref{firstpage}--\pageref{lastpage}}
\maketitle

\begin{abstract}
The debris disc around HD~172555 was recently imaged in near-infrared polarised scattered light by the Very Large Telescope's Spectro-Polarimetric High-contrast Exoplanet REsearch instrument.
Here we present optical aperture polarisation measurements of HD~172555 by the HIgh Precision Polarimetric Instrument (HIPPI), and its successor HIPPI-2 on the Anglo-Australian Telescope. We seek to refine constraints on the disc's constituent dust grains by combining our polarimetric measurements with available infrared and millimetre photometry to model the scattered light and continuum emission from the disc.
We model the disc using the 3D radiative transfer code {\sc Hyperion}, assuming the orientation and extent of the disc as obtained from the SPHERE observation. 
After correction for the interstellar medium contribution, our multi-wavelength HIPPI/-2 observations (both magnitude and orientation) are consistent with the recent SPHERE polarisation measurement with a fractional polarisation $p = 62.4~\pm~5.2$~ppm at 722.3~nm, and a position angle $\theta = 67~\pm~3\degr$. The multi-wavelength polarisation can be adequately replicated by compact, spherical dust grains (i.e. from Mie theory) that are around 1.2~$\mu$m in size, assuming astronomical silicate composition, or 3.9~$\mu$m assuming a composition derived from radiative transfer modelling of the disc. We were thus able to reproduce both the spatially resolved disc emission and polarisation with a single grain composition model and size distribution.
\end{abstract}




\section{Introduction}

HD 172555 is a young, A7~V type star with a bright, warm debris disc \citep[$L_{\rm dust}/L_{\star}~=~7.72\times10^{-4}$ , $T_{\rm dust} = 285$~K, ][]{2016Cotten}. As a member of the $\beta$ Pictoris moving group, the system has a well-determined age of 23~$\pm$~3~Myr \citep{2014MamajekBell}. The inclination and orientation of the debris disc around HD~172555 were first constrained at mid-infrared wavelengths through combining MIDI interferometric measurements with reanalysis of mid-infrared continuum observations taken by the TReCS instrument \citep{2010Moerchen,2012Smith}. Recently, \citet{2018Engler} directly imaged the disc in scattered light using SPHERE/ZIMPOL polarimetry. They found the high inclination orientation ($i = 103.5$\degr, $\phi = 112.3$\degr) and measured an outer extent of 8.5 to 11.3~au for the disc. They also determined a lower limit to the polarisation of 62~$\pm$~6~ppm in the VBB filter ($\lambda_{\rm c}$ = 735~nm, $\Delta\lambda$ = 290~nm). 

The substantial mid-infrared continuum emission observed by \textit{Spitzer} was identified as being anomalous due its brightness, and inconsistent with a steady-state origin for the circumstellar debris \citep{2007Wyatt}. Unusually for a debris disc, the \textit{Spitzer}/IRS mid-infrared spectrum of HD~172555 shows features, including SiO gas, consistent with the dust grains originating in a high velocity impact \citep{2009Lisse,2012Johnson}. Based on infrared spectroscopy, the dust composition is inferred to be differentiated silicate material similar to Asterodial bodies in the Solar system \citep{2012Morlok}. High-resolution stellar spectroscopy has also revealed variable absorption in ultra-violet and optical lines, interpreted as evidence of cometary activity in the system \citep{2014Kiefer,2018Grady}. Multi-epoch mid-infrared spectroscopic observations of HD~172555 show no substantial change ($< 20~$per cent) in the shape of the 10~$\mu$m silicate feature, suggesting the sub-micron dust grain population is stable on decadal timescales \citep{2020Su}. At far-infrared wavelengths, \textit{Herschel} photometric and spectroscopic observations revealed the presence of gas in the disc through detection of 63~$\mu$m [OI] emission, but no evidence of a second cool dust component in the system \citep{2012RiviereMarichalar,2014RiviereMarichalar}. Overall, these observations present a picture of a system in a state of dynamic disarray undergoing some collisional violence.

Here we present multi-wavelength optical aperture polarimetric measurements of HD~172555 taken with the HIgh Precision Polarimetric Instrument \citep[HIPPI][]{2015Bailey}, and its successor HIPPI-2 \citep{2020Bailey}, on the 3.9-m Anglo-Australian Telescope (AAT). These observations record the scattering-induced polarisation signature of debris dust around HD~172555. Combining these measurements with the architecture, as determined by the SPHERE/ZIMPOL observations, and ancillary mid- and far-infrared photometry measurements tracing the thermal emission from the dust, we seek to better constrain the size and albedo of the disc's constituent dust grains \citep[following e.g.][]{2018Choquet,2018Marshall}. The results of our analysis are compared to the values expected for dust grains present in the disc inferred from the presence of spectral features in the mid-infrared spectrum.

The remainder of the paper is laid out as follows. We present the multi-wavelength HIPPI/-2 polarimetric observations in Section \ref{sec:obs}. The impact of polarisation induced by dust in the interstellar medium is dealt with in Section \ref{sec:ism}. We combine the ISM-subtracted polarimetric measurements with archival photometric data to produce a self-consistent model of the disc's scattered light and continuum emission in Section \ref{sec:mod}. A comparison of our results with the dust composition inferred from the mid-infrared spectrum is given in Section \ref{sec:dis}. Finally, we present our conclusions in Section \ref{sec:con}.

\section{Observations}
\label{sec:obs}

\subsection{Ancilliary data}

The stellar parameters used in this work, including the luminosity, effective temperature, and radius, were taken from \cite{2018Gaia}. Photometry spanning infrared wavelengths have been assembled from various literature sources including \textit{IRAS} \citep{1984Neugebauer}, \textit{Akari} \citep{2010Ishihara}, \textit{WISE} \citep{2010Wright}, \textit{Spitzer} \citep{2006Chen,2014Chen}, and \textit{Herschel} \citep{2014RiviereMarichalar}. We also determine an upper limit to the millimetre flux of the disc based on a non-detection of the disc (and star) in an archival ALMA Band 6 imaging observation(project 2013.1.01147.S, P.I. S. Perez), using the pipeline reduced imaging data obtained from the Japanese Virtual Observatory\footnote{\href{http://jvo.nao.ac.jp/portal/}{Japanese Virtual Observatory}}.

\begin{table}
\centering
\tabcolsep 12.5 pt
\caption{Stellar parameters. \label{tab:star}}
\begin{tabular}{lcc}
    \hline\hline
    Parameter & Value & Ref. \\
    \hline
    Right Ascension ({\it hms}) & 18 45 26.90       & 1 \\
    Declination ({\it dms})     & -64 52 16.5       & 1 \\
    Proper motions (mas/yr)     & 32.073, -150.182  & 1 \\
    Distance (pc)               & 28.34~$\pm$~0.18  & 1 \\
    $V$ (mag)                   & 7.513~$\pm$~0.005 & 2 \\
    $B-V$ (mag)                 & 0.200~$\pm$~0.016 & 2 \\
    Spectral type               & A7~V              & 3 \\
    Luminosity ($L_{\odot}$)    & 8.093$~\pm~$0.062 & 1 \\
    Radius ($R_{\odot}$)        & 1.55$^{+0.17}_{-0.02}$    & 1 \\
    Temperature (K)             & 7816$^{+60}_{-397}$       & 1 \\
    Surface gravity, $\log g$   & 4.18 & 3 \\
    Metallicity, [Fe/H]         & 0.09 & 3 \\
    $\nu \sin i$ (km/s)         & 2.0~$\pm$~4.2 & 3 \\
    Age (Myr)                   & 23~$\pm$~2 & 4 \\
    \hline
\multicolumn{3}{l}{\textbf{References:} 1. \cite{2018Gaia},}\\
\multicolumn{3}{l}{2. \citep{2000Hog}, 3. \citep{2006Gontcharov},}\\
\multicolumn{3}{l}{4. \citep{2014MamajekBell} }\\
\end{tabular}
\end{table}

\subsection{HIPPI/-2 aperture polarimetry}

We obtained multi-wavelength optical and near-infrared aperture polarimetry observations of HD~172555 and a number of interstellar calibration stars using the HIgh Precision Polarimetric Instrument \citep[HIPPI;][]{2015Bailey} and its successor\footnote{HIPPI-2 has been developed based on experience with HIPPI and Mini-HIPPI \citep{2017Bailey}.} \citep[HIPPI-2;][]{2020Bailey} on the 3.9-m Anglo-Australian Telescope. HIPPI has been successfully used for a range of science programs including surveys of polarisation in bright stars \citep{2016Cotton}, the first detection of polarisation due to rapid rotation in hot stars \citep{2017aCotton}; reflection from the photospheres of binary stars \citep{2019Bailey} and the most sensitive searches for similar effects from exoplanets \citep{2016Bott,2018Bott}; studies of the polarisation in active dwarfs \citep{2017bCotton,2019aCotton}, debris discs \citep{2017bCotton}, the interstellar medium \citep{2017bCotton,2019bCotton} and hot dust \citep{2016Marshall}. HIPPI-2 has recently been used in the study of reflected light in binary systems \citep{2019Bailey,2020cCotton}, the rapidly rotating system $\alpha$~Oph \citep{2020bBailey}, the red supergiant Betelgeuse \citep{2020aCotton} and the polluted white dwarf G29-38 \citep{2020bCotton}.

Mounted at the AAT f/8 Cassegrain focus, the HIPPI aperture has a diameter of 6.6\arcsec, such that the disc of HD~172555 lies fully within it (With HIPPI-2 an aperture corresponding to 11.9\arcsec was selected). The observations were taken over four observing runs spanning May 24th, 2015 to September 1st, 2018.

HIPPI-class polarimeters achieve very high precision through the use of a Ferro-electric Liquid Crystal (FLC) modulator operating at 500 Hz to beat seeing noise. On the AAT HIPPI-2 demonstrates a wavelength dependant ultimate precision, ranging from $\sim$1 part-per-million (ppm) at red wavelengths, $\sim$ 2 ppm at green wavelengths, and $\sim$13 ppm for a passband with an effective wavelength of 400 nm; with HIPPI-2 typically achieving a precision 1 ppm better than HIPPI in any given band \citep{2020Bailey}.

We made observations in six different filter bands: two short-pass filters below 425 nm (425SP) and 500 nm (500SP) respectively, SDSS \gfil~and \rfil\footnote{Two different versions of the \gfil~and \rfil~filters were used; the Astrodon generation 2 versions were used with HIPPI-2.}, a Johnson V filter, and a long-pass filter above 650~nm (650LP). These were paired with either blue-sensitive (B) Hamamatsu H10720-210 modules which have ultrabialkali photocathodes or red sensitive (R) Hamamatsu H10720-20 modules with extended red multialkali photocathodes as detectors. The same MS Series polarisation rotator from Boulder Non-linear Systems was used as a modulator for all of the reported observations with both versions of the instrument. The modulator's performance has drifted over time, from having a wavelength of peak efficiency of 494.8 nm in 2014 to 595.4 nm at the end of the August 2018 run. Our use of the modulator has therefore been broken down into performance eras, the characteristics of which are described in \citet{2020Bailey} and the supplementary materials of \citet{2019Bailey}. The bandpass model uses the characteristics of the optical components, the airmass and source spectrum, to calculate the band effective wavelength ($\lambda_{\rm eff}$) and modulation efficiency (Eff.) for each observation. A raw observation is multiplied by the inverse of Eff. to give the true polarisation.

An observation with HIPPI/-2 consists of four measurements of the target at instrumental position angles of 0\degr, 45\degr, 90\degr and 135\degr, with an accompanying sky measurement also made at each position angle. The sky background is first subtracted, and then the measurements combined to eliminate instrumental polarisation. The observations are then rotated from the instrumental frame to the sky frame by reference to measurements of high polarisation standard stars. The standards, of which details are given in \citet{2020Bailey}, have uncertainties of about a degree. These high polarisation standard observations are made in either \gfil~or without a filter. However, because of drift in the modulator performance, second order corrections of 5.8\degr and 2.6\degr were needed for the 500SP and 425SP filter measurements for the 2018AUG run, based on multi-band standard observations. A small wavelength dependant polarisation is imparted by the telescope (TP), which must be determined by the measurement of unpolarised standard stars, and then subtracted. The TP is determined as the mean of all standards observed in a given band, equally weighted. Thence the error in the TP is incorporated into the errors in the measurements of science targets as the square root of the sum of the errors squared. On occasion the TP has been stable across consecutive runs, and in such cases observations from both are included in the adopted TPs. A summary of standard observations relevant to the observations made here and some additional details are presented in Appendix \ref{sec:std_obs}.

\begin{table*}
\caption{HIPPI/-2 polarisation measurements of HD~172555 and its interstellar calibrators. \label{tab:pol}}
\tabcolsep 4.5 pt
\begin{tabular}{lcccrcccccrr}
    \hline\hline
    Target    & UT & Run & Instr. & Ap. & Mod. & Fil & PMT & $\lambda_{\rm eff}$ & Eff. & \multicolumn{1}{c}{$q$} & \multicolumn{1}{c}{$u$} \\
              & & & & (\arcsec) & Era & & & (nm) & & \multicolumn{1}{c}{(ppm)} & \multicolumn{1}{c}{(ppm)}    \\
    \hline
HD 172555 & \02015-05-25 04:47:17* & 2015MAY & HIPPI & 6.6 & E1 & 425SP & B & 400.4 & 0.569 & -17.0~$\pm$~25.6 & 154.3~$\pm$~26.0 \\
HD 172555 & 2015-05-25 15:14:49 & 2015MAY & HIPPI & 6.6 & E1 & 425SP & B & 400.2 & 0.568 & -68.9~$\pm$~20.0 & 110.8~$\pm$~19.7 \\
HD 172555 & 2018-08-27 11:27:42 & 2018AUG & HIPPI-2 & 11.9 & E6 & 500SP & B & 440.4 & 0.479 & -103.9~$\pm$~\07.9 & 106.2~$\pm$~\08.2 \\
HD 172555 & 2018-08-27 11:59:25 & 2018AUG & HIPPI-2 & 11.9 & E6 & 500SP & B & 440.4 & 0.479 & -105.5~$\pm$~\07.6 & 112.6~$\pm$~\07.9 \\
HD 172555 & 2018-08-27 12:29:52 & 2018AUG & HIPPI-2 & 11.9 & E6 & \gfil& B & 466.1 & 0.643 & -96.5~$\pm$~\06.0 & 117.6~$\pm$~\06.0 \\
HD 172555 & 2015-05-24 17:53:44 & 2015MAY & HIPPI & 6.6 & E1 & \gfil& B & 467.2 & 0.899 & -111.5~$\pm$~\07.0 & 110.6~$\pm$~\07.1 \\
HD 172555 & 2018-08-28 14:39:16 & 2018AUG & HIPPI-2 & 11.9 & E7 & V & B & 534.3 & 0.898 & -103.8~$\pm$~\08.6 & 91.7~$\pm$~\08.1 \\
HD 172555 & 2018-08-28 15:06:53 & 2018AUG & HIPPI-2 & 11.9 & E7 & V & B & 534.4 & 0.898 & -110.5~$\pm$~\08.0 & 94.3~$\pm$~\08.3 \\
HD 172555 & 2015-05-24 18:33:20 & 2015MAY & HIPPI & 6.6 & E1 & \rfil & B & 599.8 & 0.833 & -125.2~$\pm$~19.2 & 110.5~$\pm$~18.7 \\
HD 172555 & 2018-08-26 13:09:54 & 2018AUG & HIPPI-2 & 11.9 & E6 & \rfil& R & 623.2 & 0.922 & -103.7~$\pm$~\06.2 & 86.6~$\pm$~\06.3 \\
HD 172555 & 2018-08-26 13:36:58 & 2018AUG & HIPPI-2 & 11.9 & E6 & \rfil& R & 623.3 & 0.922 & -102.7~$\pm$~\06.1 & 93.0~$\pm$~\05.8 \\
HD 172555 & 2018-08-23 10:37:43 & 2018AUG & HIPPI-2 & 11.9 & E5 & 650LP & R & 722.3 & 0.738 & -97.6~$\pm$~10.6 & 91.2~$\pm$~10.5 \\
HD 172555 & 2018-08-24 11:47:30 & 2018AUG & HIPPI-2 & 11.9 & E6 & 650LP & R & 722.3 & 0.775 & -87.9~$\pm$~10.2 & 83.6~$\pm$~10.6 \\
HD 172555 & 2018-08-26 12:15:13 & 2018AUG & HIPPI-2 & 11.9 & E6 & 650LP & R & 722.3 & 0.775 & -109.7~$\pm$~10.1 & 91.5~$\pm$~10.4 \\
HD 172555 & 2018-08-26 12:42:00 & 2018AUG & HIPPI-2 & 11.9 & E6 & 650LP & R & 722.3 & 0.775 & -116.4~$\pm$~10.3 & 67.8~$\pm$~\09.5 \\
\hline
HD 162521 & 2018-08-29 10:15:34 & 2018AUG & HIPPI-2 & 11.9 & E7 & 500SP & B & 443.8 & 0.457 & -93.0~$\pm$~18.2 & 73.9~$\pm$~18.2 \\
HD 162521 & 2018-08-29 10:46:09 & 2018AUG & HIPPI-2 & 11.9 & E7 & 500SP & B & 443.8 & 0.457 & -59.3~$\pm$~17.9 & 67.8~$\pm$~18.3 \\
HD 162521 & 2018-08-21 15:22:52 & 2018AUG & HIPPI-2 & 11.9 & E5 & \gfil& B & 470.3 & 0.737 & -106.7~$\pm$~16.7 & 83.3~$\pm$~15.9 \\
HD 162521 & 2018-08-29 11:16:41 & 2018AUG & HIPPI-2 & 11.9 & E7 & \gfil& B & 469.0 & 0.608 & -94.5~$\pm$~13.1 & 48.5~$\pm$~12.6 \\
HD 162521 & 2018-08-29 11:45:44 & 2018AUG & HIPPI-2 & 11.9 & E7 & V & B & 535.7 & 0.901 & -95.5~$\pm$~18.1 & 73.5~$\pm$~18.2 \\
HD 162521 & 2018-08-29 12:13:53 & 2018AUG & HIPPI-2 & 11.9 & E7 & V & B & 535.7 & 0.901 & -88.5~$\pm$~17.2 & 62.5~$\pm$~18.3 \\
HD 162521 & 2018-09-01 14:16:39 & 2018AUG & HIPPI-2 & 11.9 & E7 & 650LP & R & 724.1 & 0.860 & -91.3~$\pm$~26.2 & -2.1~$\pm$~24.0 \\
HD 162521 & 2018-09-01 14:34:07 & 2018AUG & HIPPI-2 & 11.9 & E7 & 650LP & R & 724.1 & 0.861 & -25.5~$\pm$~27.5 & 63.1~$\pm$~25.2 \\
HD 162521 & 2018-09-01 14:56:18 & 2018AUG & HIPPI-2 & 11.9 & E7 & 650LP & R & 724.1 & 0.861 & -83.7~$\pm$~18.2 & 40.7~$\pm$~19.3 \\
HD 165499 & 2018-08-22 15:53:00 & 2018AUG & HIPPI-2 & 11.9 & E5 & \gfil& B & 472.4 & 0.748 & 33.2~$\pm$~11.7 & 23.6~$\pm$~11.6 \\
HD 167425 & 2018-08-27 13:52:13 & 2018AUG & HIPPI-2 & 11.9 & E6 & 500SP & B & 446.4 & 0.522 & -105.9~$\pm$~16.4 & 72.6~$\pm$~16.4 \\
HD 167425 & 2018-08-27 14:21:50 & 2018AUG & HIPPI-2 & 11.9 & E6 & 500SP & B & 446.6 & 0.523 & -85.1~$\pm$~16.2 & 113.6~$\pm$~16.1 \\
HD 167425 & 2018-08-22 15:32:12 & 2018AUG & HIPPI-2 & 11.9 & E5 & \gfil& B & 471.7 & 0.744 & -115.0~$\pm$~15.3 & 80.3~$\pm$~12.7 \\
HD 167425 & 2018-08-27 14:50:42 & 2018AUG & HIPPI-2 & 11.9 & E6 & \gfil& B & 471.5 & 0.677 & -104.6~$\pm$~13.2 & 68.0~$\pm$~13.7 \\
HD 167425 & 2018-08-24 13:42:16 & 2018AUG & HIPPI-2 & 11.9 & E6 & \rfil& R & 625.5 & 0.920 & -90.8~$\pm$~12.6 & 62.6~$\pm$~12.6 \\
HD 167425 & 2018-08-26 14:32:21 & 2018AUG & HIPPI-2 & 11.9 & E6 & \rfil& R & 625.6 & 0.920 & -136.9~$\pm$~13.7 & 82.6~$\pm$~13.1 \\
HD 167425 & 2018-08-24 13:17:05 & 2018AUG & HIPPI-2 & 11.9 & E6 & 650LP & R & 725.1 & 0.766 & -102.9~$\pm$~17.5 & 79.7~$\pm$~16.6 \\
HD 167425 & 2018-08-26 14:06:48 & 2018AUG & HIPPI-2 & 11.9 & E6 & 650LP & R & 725.1 & 0.766 & -79.3~$\pm$~18.3 & 72.6~$\pm$~16.6 \\
HD 173168 & 2017-06-25 13:25:23 & 2017JUN & HIPPI & 6.6 & E2 & \gfil& B & 467.5 & 0.875 & -153.1~$\pm$~\09.1 & 159.2~$\pm$~\09.0 \\
HD 177389 & 2018-08-21 15:01:19 & 2018AUG & HIPPI-2 & 11.9 & E5 & \gfil& B & 474.8 & 0.761 & 4.9~$\pm$~10.5 & -12.9~$\pm$~\09.8 \\
HD 186219 & 2015-10-20 10:37:17 & 2015OCT & HIPPI & 6.6 & E1 & \gfil& B & 466.6 & 0.898 & 78.3~$\pm$~\08.3 & 22.8~$\pm$~\08.3 \\
    \hline
\end{tabular}
*Half of this observation was made on each of two consecutive nights at similar airmass, 2019-05-24 and 2019-05-25. 
\end{table*}

\begin{table}
\tabcolsep 7 pt
\caption{Interstellar calibrators for HD~172555. \label{tab:ism}}
\begin{tabular}{lcccc}
    \hline\hline
    Calibrator    &   V    & Sp. Type &    $d$   &   Sep     \\
                  & (mag)  &          &   (pc)   & (\degr)   \\
    \hline
    HD~162521     &  6.36  & F5~V     &  35.52~$\pm$~0.05  &   \05.2   \\
    HD~165499     &  5.47  & G0~V     &  17.75~$\pm$~0.04  &   \04.8   \\
    HD~167425     &  6.17  & F9.5~V   &  23.05~$\pm$~0.21  &   \03.0   \\
    HD~173168     &  5.70  & A8~V     &  65.25~$\pm$~0.31  &   \05.0   \\
    HD~177389     &  5.31  & K0~IV    &  36.98~$\pm$~0.12  &   \04.3   \\
    HD~186219     &  5.39  & A4~IV    &  43.42~$\pm$~0.27  &   \09.5   \\
    \hline
    HD~2151       &  2.79  & G0~V     & \07.46~$\pm$~0.01  &    27.0   \\
    HD~156384     &  5.89  & K3~V+K5~V& \06.84~$\pm$~0.40  &    32.6   \\
    HD~165135     &  2.99  & K0~III   &  29.70~$\pm$~0.16  &    35.0   \\
    HD~209100     &  4.69  & K5~V     & \03.64~$\pm$~0.00  &    24.7   \\
    \hline
\end{tabular}
\end{table}

Observations (table \ref{tab:pol}) were made of both HD~172555 and a number of other nearby stars to investigate the properties of the interstellar medium. These interstellar control stars, listed above the line in table \ref{tab:ism}\footnote{The stars listed below the line are those from the \textit{Interstellar List} in \citet{2017bCotton} that are within 35\degr angular separation.}, were selected on the basis that they lay within 10\degr~of HD~172555 on the sky, had approximately the same distance ($\delta d \sim 10$~per cent), exhibited no evidence of infrared excess (through cross matching with infrared catalogues), and were of spectral types unlikely to exhibit intrinsic polarisation \citep{2016Cotton,2016bCotton,2017bCotton}. Distances were taken from \cite{2018Gaia}, spectral types from \cite{2006Gray}, and $V$ magnitudes from \cite{2000Hog}.

The summary of the HIPPI/-2 polarisation observations of HD~172555 and the interstellar calibrators given in Table \ref{tab:pol} present the data in the form of normalised Stokes parameters $q=Q/I$ and $u=U/I$, along with the internal error of the observations ($\sigma_o$). To obtain the true error, $\sigma$, the instrumental precision in the corresponding band, $\sigma_i$, needs to be added as $\sigma=\sqrt{\sigma_o^2+\sigma_i^2}$. The polarisation, $p$, can be calculated as $p=\sqrt{q^2+u^2}$ and the polarisation position angle, $\theta$, as $\theta=\frac{1}{2}\tan^{-1}\left (\frac{u}{q}  \right )$. When considering $p$, since it is positive definite, it is common to debias it according to \begin{equation}\hat{p}=\begin{cases}\sqrt{p^2-\sigma_p^2} & \text{ for } p>\sigma_p \\  0 & \text{ for } p<\sigma_p\end{cases}.\end{equation}

\section{Interstellar polarisation}
\label{sec:ism}
\subsection{Polarisation vs. distance}

\begin{figure*}
    \centering
    \includegraphics[width=\textwidth, trim={2cm 0.5cm 2cm 1.5cm},clip]{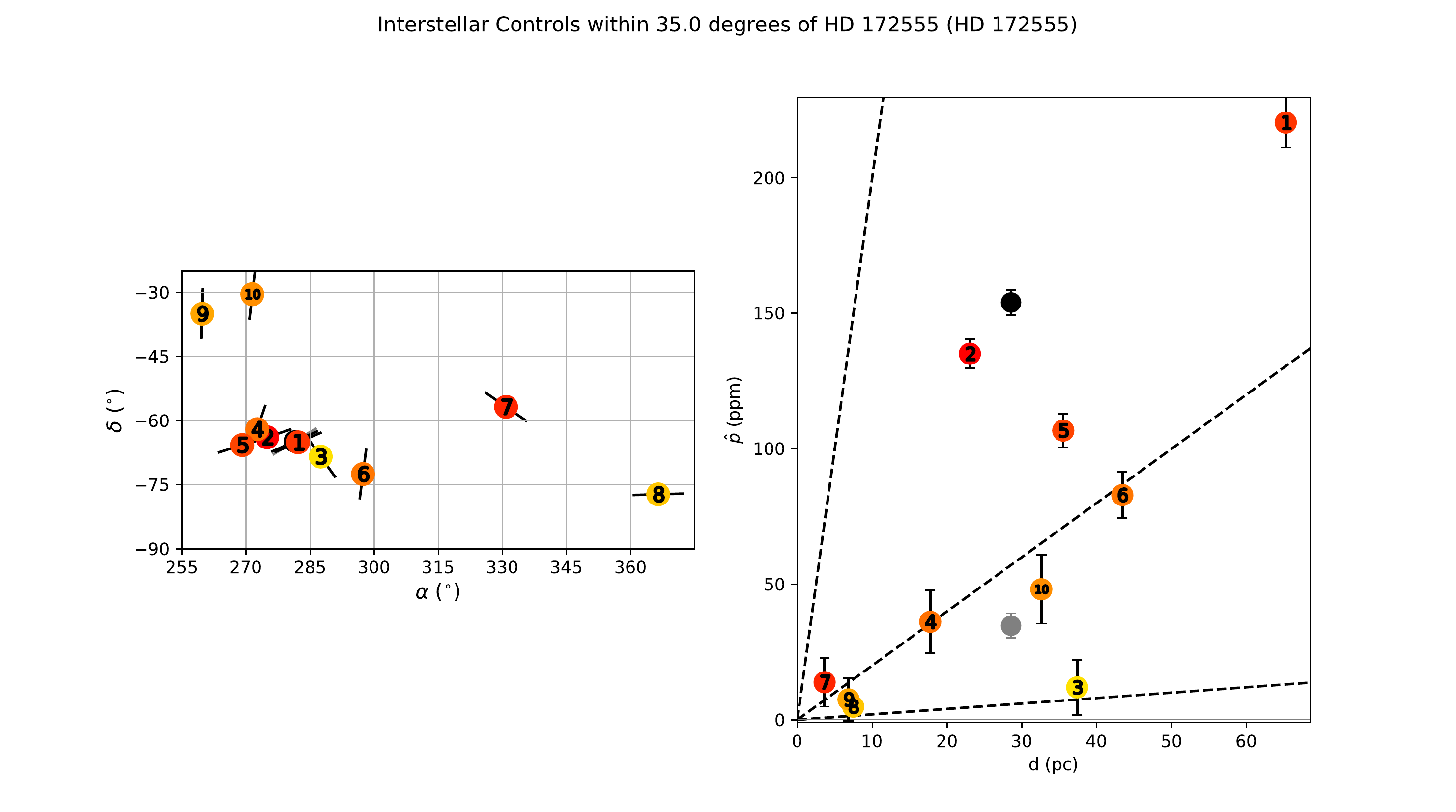}
    \caption{\textit{Left} Plot illustrating polarisation as a function of sky position. Unintrinsically polarised stars within 35\degr of HD~172555 (black) are shown as circles with their position angles (clockwise from North to East) as bars; their colours from red to yellow are encoded based on $\hat{p}/d$, with the numbers indicating their order in terms of separation. The model $\theta$ as calculated from \citet{2017bCotton} for HD~172555 is shown in grey. The stars are: 1 HD~173168, 2: HD~167425, 3: HD~177389, 4: HD~165499, 5: HD~162521, 6: HD~186219, 7: HD~209100, 8: HD~2151, 9: HD~156384, HD~165135. \textit{Right}: Plot illustrating (debiased) polarisation as a function of distance. The object colours and numbers are as per the left panel, with the grey circle indicating the model interstellar polarisation calculated as in \citet{2017bCotton}. The three dashed lines show polarisation increasing at 0.2, 2.0 and 20 ppm/pc. \label{fig:ism}}
\end{figure*}

An accurate interstellar subtraction is critical for investigating the polarisation properties of a debris disc system. In the case of HD~172555 this task is made easier by the existence of imaging data that can separate the polarisation of the disc and the star -- allowing the interstellar polarisation to be found. The imaging data is monochromatic though, so the wavelength dependence of the interstellar polarisation still needs to be determined in order to carry out an interstellar subtraction in other bands.

Fig. \ref{fig:ism} shows the polarisation in \gfil~of the interstellar controls (and a few intrinsically unpolarised stars from \citet{2017bCotton}) compared to HD~172555. HD~172555 is more polarised in terms of polarisaiton vs distance ($p/d$) than any of the nearby stars. The two closest stars (in terms of angular separation, HD~173168 and HD~167425) are more polarised than most of those near the Sun seen in earlier work \citep{2010Bailey,2016Cotton,2016Marshall,2017bCotton}, being more than double the $p/d$ predicted by the model in \citet{2017bCotton} for stars within 25~pc. These two stars, and HD~162521 -- which is similarly polarised -- have very similar polarisation angles to HD~172555. Other nearby stars, in particular HD~177389, are not so polarised showing the ISM in this region is patchy which prevents firm conclusions being drawn on the interstellar polarisation of HD~172555 from this data alone. On balance though, it suggests a large fraction of the polarisation is interstellar.

\subsection{Wavelength dependence}

The wavelength dependence of interstellar (linear) polarisation is given by the empirically determined Serkowski Law \citep{1975Serkowski} as: \begin{equation}\label{eq:Serk} \frac{p(\lambda)}{p_{\rm max}}=exp\left[-Kln^2\left (\frac{\lambda_{\rm max}}{\lambda} \right )  \right ] ,\end{equation} where $p(\lambda)$ is the polarisation at wavelength $\lambda$, $p_{\rm max}$ is the maximum polarisation occurring at wavelength $\lambda_{\rm max}$. The dimensionless constant $K$ describes the inverse width of the polarisation curve peaked around $\lambda_{\rm max}$; \citet{1975Serkowski} gave its value as 1.15. \citet{1980wilking} later described $K$ in terms of a linear function of $\lambda_{\rm max}$. Using this form, \citet{1992Whittet} find $K$ to be: \begin{equation}\label{eq:Whit} K=(0.01\pm0.05)+(1.66\pm0.09)\lambda_{\rm max},\end{equation} (where $\lambda_{\rm max}$ is given in $\mu$m) -- the form of equation \ref{eq:Serk} that uses this relation is referred to as the Serkowski-Wilking Law. There have been only three studies which have determined $\lambda_{\rm max}$ for stars near to the Sun. In the most recent multi-object study, \citet{2019bCotton} found 350 $\pm$ 50 nm for four stars within the Local Hot Bubble, and 550 $\pm$ 20 nm for two stars in its wall. Earlier \citet{2016Marshall} found $\lambda_{\rm max}$ to be between 35 nm and 600 nm with a most probable value of 470 nm from 2-band observations of a few stars at Southern declinations. Lastly, in investigating the rotational properties of $\alpha$~Oph, \citet{2020bBailey} found 440 $\pm$ 110 nm for that star. 

\begin{figure}
\includegraphics[width=\columnwidth, trim={0.5cm 0.5cm 0.0cm 0.5cm}]{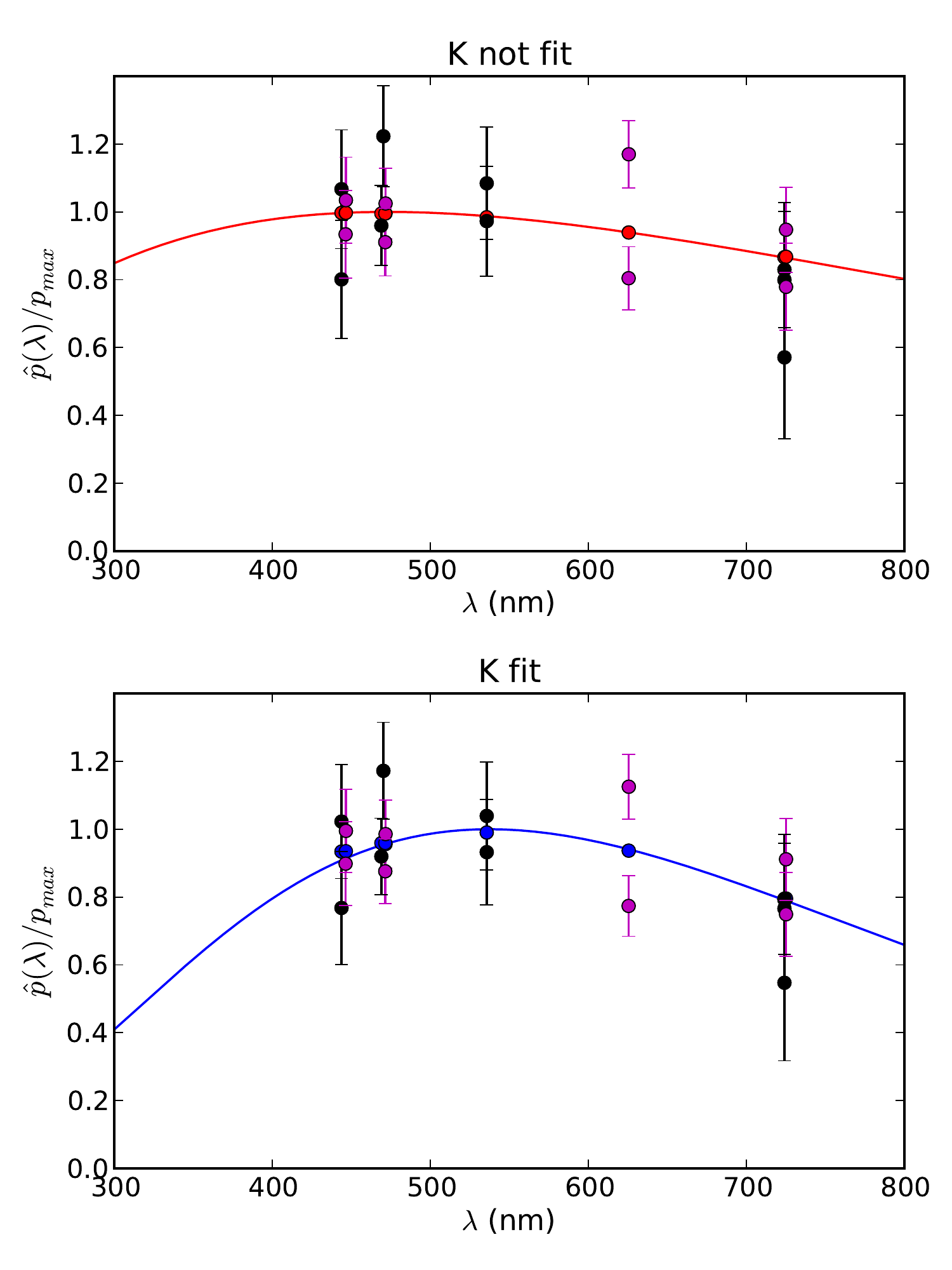}
\caption{The results of Serkowski fits to two targets -- HD~162521 (black), HD~167425 (magenta). In both panels $p_{\rm max}$ is fit for each target, along with $\lambda_{\rm max}$, in the lower panel $K$ is also fit. The data points in black and magenta are shown as fractional polarisation, $\hat{p}/p_{\rm max}$ for the target, and placed at the effective wavelength $\lambda_{\rm eff}$ for the band. The fit Serkowski curves are shown in red (without $K$) and blue (with $K$); the corresponding coloured points correspond to the fits for the bands corresponding to the data points.}
\label{fig:comb_Serk}
\end{figure}

\begin{table}
\tabcolsep 2.6 pt
\centering
\caption{Serkowski fits to interstellar polarisation calibrator stars.}
\label{tab:comb_Serk}
\begin{tabular}{cccclc}
\hline
\hline
Fit $K$ &   \multicolumn{2}{c}{$p_{\rm max}$}                   &   $\lambda_{\rm max}$  &   \0\0\0$K$       &   $\chi^2_r$  \\
        &   \multicolumn{2}{c}{(ppm)}                       & (nm)               &                   &               \\
        &   HD~162521            &   HD~167425      \\
\hline
No      &   109.9 $\pm$ 7.1       &   148.9 $\pm$ \07.4      &   472.6 $\pm$ 79.0 &  0.79*            &   1.10        \\
Yes     &   114.6 $\pm$ 9.8       &   141.6 $\pm$ 10.3       &   537.2 $\pm$ 33.8 &  2.63 $\pm$ 2.12  &   1.12        \\
\hline
\end{tabular}
\begin{flushleft}
* Calculated using equation \ref{eq:Whit}. \\
\end{flushleft}	
\end{table}

Multi-band data was taken for HD~167425 and HD~162521. As stars with similar $p/d$, $\theta$, $d$ and small separations from HD~172555, they were chosen as proxies to investigate its interstellar polarisation wavelength dependence. The same methods and procedure as followed in \citet{2019bCotton} were adopted for this purpose. Using a full bandpass model we use \textsc{Python's} \textsc{curve\_fit} program to fit both the Serkowski-Wilking Law ($K$ not fit) and the Serkowski Law ($K$ fit). The method fits $p_{\rm max}$ for both control stars, along with $\lambda_{\rm max}$, and in the latter case $K$ as well. The results are graphed in Fig. \ref{fig:comb_Serk}, and the fit values reported in table \ref{tab:comb_Serk}.

The Serkowski and Serkwoski-Wilking Law fits are similarly good, as judged by the reduced $\chi^2$ statistic, and the values of each fit parameter agree within the stated uncertainties. The values of $p_{\rm max}$, and hence $p/d$ are higher than was expected for this part of the interstellar medium, and may indicate a transition region between the bulk Local Hot Bubble and its wall (insomuch as these regions are useful as concepts). The Serkowski Law gives a redder, more tightly defined $\lambda_{\rm max}$, but $K$ is not particularly well defined. The past studies of the interstellar medium \citep{2016Marshall,2019bCotton,2020bBailey} have found values of $\lambda_{\rm max}$ more closely aligned with the Serkowski-Wilking Law fit, and so values of $\lambda_{\rm max}$ equal to 472.6 nm and $K$ equal to 0.79 are adopted.

\subsection{Interstellar subtraction}

To remove the interstellar polarisation we make the assumptions that the interstellar polarisation is described by the Serkowski-Wilking Law, has a consistent orientation with wavelength, and that the intrinsic polarisation of the star-disc system has a consistent orientation with wavelength too. That is, if the interstellar polarisation is removed the remaining polarisation can be rotated such that $u^{\prime}$ is 0 in all bands. We consider the 650LP observations to be equivalent to the SPHERE VBB band observations given their similar $\lambda_{\rm eff}$, 725 nm and 735 nm respectively, so in this case $q^{\prime}$ should also match the measured integrated disc$+$star polarisation reported by \citet{2018Engler} of 62 $\pm$ 8 ppm.

To carry out the subtraction $\theta_i$, $p_{\rm max}$ and $\theta_{\star}$ -- where the subscripts $i$ and $\star$ represent respectively interstellar and intrinsic -- need to be determined. This is done using a \textsc{Python} \textsc{curve\_fit}-based program which tries to minimise the difference between a set of model values and the data as modified by a function. 

Initial values for the program are: $p_{\rm max}=135.4$ ppm and $\theta_i=59.5\degr$, chosen based on, respectively, an average of the $p/d$ of HD~161521 and HD~167425 and the interstellar model in \citet{2017bCotton}; and $\theta_{\star}=22.3\degr$ based on $90\degr$ minus the disc position angle\footnote{For small grains and a symmetric disc, the polarisation is expected to be perpendicular to the major axis.} of 112.3\degr $\pm$ 1.5 found by \citet{2018Engler}. The \textsc{curve\_fit} program then carries out the following steps with each iteration of fit parameters: 
\begin{itemize}
    \item Takes the raw observational data and subtracts $q_i$ and $u_i$ from each band, based on the bandpass calculation of the Serkowski-Wilking Law described interstellar polarisation with $\lambda_{\rm max}$ of 472.6 nm, $K$ of 0.79, and $p_{\rm max}$ and $\theta_i$ chosen by that \textsc{curve\_fit} iteration.
    \item Rotates the remaining (intrinsic) polarisation by $\theta_{\star}$ into a new frame ($q^{\prime}$, $u^{\prime}$).
    \item Returns $u^{\prime}$ for each observation, and for the 650LP observations also $62-q^{\prime}$, for comparison with model values of 0 (the $q^{\prime}$ errors are modified to incorporate the uncertainty in the \citet{2018Engler} measurement).
\end{itemize}

Table \ref{tab:isfit} summarises the fit results, whereas table \ref{tab:indisc} presents band averaged (error weighted) values for the intrinsic polarisation of the system, i.e. that remaining after the interstellar subtraction. The result is a $p/d$ similar to that of HD~167425, a value of $\theta_{\star}$ consistent with that of \citet{2018Engler}, and values of $u^{\prime}$ consistent with zero to close to 1-$\sigma$ or better in each band, which all give us confidence in the determination. 

A fit allowing $\lambda_{\rm max}$ to be free was also tried, but in that case the value shifts to $\sim$400 nm -- equivalent to the 425SP $\lambda_{\rm eff}$ -- in order to eliminate the residual in $u^{\prime}$ in the shortest wavelength band, in so doing the value of $p_{\rm max}$ is pushed up to less realistic levels and $\theta_{\star}$ is reduced such that it is no longer consistent with the value obtained by \citet{2018Engler}. It is prudent not to overemphasise the 425SP band, given that one of the 425SP observations is a composite of two measurements, and the associated error may be under-estimated. Consequently, we favour the fit with fixed $\lambda_{\rm max}$.

\begin{table}
\tabcolsep 7.5 pt
\centering
\caption{A summary of fit parameters for HD~172555 interstellar subtraction.}
\label{tab:isfit}
\begin{tabular}{ccccc}
\hline
\hline
$p_{\rm max}$   & $\lambda_{\rm max}$*  & $K$*  &   $\theta_i$  &   $\theta_{\star}$    \\
(ppm)       & (nm)              &       &   ($\degr$)   &   ($\degr$)           \\
\hline
176.6 $\pm$ 12.1    &   472.6   & 0.79  & 82.5 $\pm$ 1.5 & 21.5$^\dag$ $\pm$ 4.5       \\
\hline
\end{tabular}
\begin{flushleft}
* Not fit. \\
$^\dag$ This implies a disc position angle of 111.5$\degr$ $\pm$ 4.5. \\
\end{flushleft}	
\end{table}

\begin{table}
\tabcolsep 8 pt
\centering
\caption{Band averaged intrinsic polarisation for HD~172555.}
\label{tab:indisc}
\begin{tabular}{ccccrr}
\hline
\hline
\multicolumn{2}{c}{Band}& n &   $\lambda_{\rm eff}$ & \multicolumn{1}{c}{$q^{\prime}$} & \multicolumn{1}{c}{$u^{\prime}$}   \\
&                       &   &   (nm)            & \multicolumn{1}{c}{(ppm)} & \multicolumn{1}{c}{(ppm)} \\   
\hline
425SP   & B & 2             & 400.3             & 143.8\0$\pm$ 18.4   & $-$19.4 $\pm$ 18.4 \\
500SP   & B & 2             & 440.4             &\090.6\0$\pm$ \07.0   & \0\02.9 $\pm$ \07.2 \\
\gfil   & B & 2             & 466.6             &\095.4\0$\pm$ \04.7   & \0\05.6 $\pm$ \04.8 \\
V       & B & 2             & 534.4             &\076.4\0$\pm$ \05.9   &\0$-$5.8 $\pm$ \05.8 \\
\rfil   & B & 1             & 599.8             &\072.6\0$\pm$ 19.3   &  \023.7 $\pm$ 18.8 \\
\rfil   & R & 2             & 623.3             &\073.2\0$\pm$ \04.4   &\0$-$4.0 $\pm$ \04.3 \\
650LP   & R & 4             & 722.3             &\062.4*$\pm$ \05.2   & \0\01.6 $\pm$ \05.1 \\
\hline
\end{tabular}
\begin{flushleft}
* Fitted to a value of 62 $\pm$ 8 ppm based on SPHERE VBB ($\lambda_{\rm eff}=$ 735 nm) measurements. \\
\end{flushleft}	
\end{table}

\section{Modelling}
\label{sec:mod}

In this section we attempt to model the ISM-subtracted aperture polarimetry measurements from HIPPI/-2 using the constraint of the observed disc orientation and architecture from the SPHERE measurements. We begin by fitting the dust continuum emission with a debris disc of the dust spatial distribution observed in scattered light \citep{2018Engler}, and inferred grain properties derived from the mid-infrared spectrum \citep{2009Lisse,2012Johnson}. Following that, we proceed with a simple scattering model of the disc to determine the best-fit dust grain size for the multi-wavelength polarimetric measurements, using the dust optical constants consistent with the composition derived from the SED fitting to infer the grain size necessary to produce the observed polarisation spectrum. Through this two step approach we demonstrate that the polarimetric measurements can provide additional, unique information to assist in the interpretation of debris dust. 

\subsection{Spectral energy distribution}

We use the 3D Monte Carlo radiative transfer code {\sc Hyperion} \citep{2011Robitaille} to model the dust continuum emission. We again adopt the best-fit parameters for the scattered light disc imaged in \cite{2018Engler} as the parameters for the disc density distribution. We have therefore tacitly assumed that the same dust grains that produce the dust continuum emission are responsible for the scattering and polarisation. The objective of this exercise is to demonstrate that the assumed properties of the dust based on spatially unresolved spectroscopic and polarimetric observations provide an adequate fit to the data when combined with the imaged spatial extent of the disc. 

We compare the continuum emission model to a dust spectral energy distribution (SED) assembled from available photometric measurements of HD~172555 covering optical to millimetre wavelengths \citep{2000Hog,2006Skrutskie,2010Wright,2010Ishihara,2014Chen,2014RiviereMarichalar}. A summary of the measurements are provided in table \ref{tab:sed}. We also extracted photometry from the \textit{Spitzer}/IRS spectrum published in \cite{2006Chen} to constrain the dust composition in the fitting, taking 14 values evenly distributed between 6 and 34~$\mu$m.

The stellar photospheric contribution to the total emission was represented by a stellar atmosphere model from the NEXTGEN grid taken from the Spanish Virtual Observatory\footnote{\href{http://svo2.cab.inta-csic.es/theory/newov2/}{Spanish Virtual Observatory}} with an effective temperature of 7800~K, surface gravity $\log (g)$ of 4.2 (interpolated from $\log (g)$ 4.0 and 4.5), and Solar metallicity which are appropriate values for HD~172555 (see Table \ref{tab:star}). The photosphere model was scaled to provide a total luminosity of 8.093~$L_{\odot}$. 

The SED fitting process was carried out in the following manner; given that the disc spatial extent and dust size distribution are fixed, the only constraint left is to determine the dust composition that best replicates the SED. The radial dust density distribution is given by a functional form 

\begin{equation}
\label{eqn:dust_dens}
\begin{split}
\rho = \rho_{\rm 0} \times ( (R/R_{0})^{-2\alpha_{\rm in}} + (R/R_{0})^{-2\alpha_{\rm out}} )^{-0.5} \\
\times (1 + ( z /(z_{0}(R/R_{0})^{\beta}))^{2})^{-1}
\end{split}
\end{equation}

\noindent where $\rho$ is the dust density at a given position around the star, $\rho_{\rm 0}$ is a density scaling factor, $R$ is the radial distance from the star, $R_{\rm 0}$ is the radial distance of the peak in disc emission from the star, $\alpha_{\rm in}$ and $\alpha_{\rm out}$ are the exponents for the slope of the disc interior and exterior to the peak position, $z$ is the vertical distance from the disc mid-plane, $z_{0}$ is the value of $z$ at $R_{0}$, and $\beta$ is the flaring index of the disc. Following \cite{2018Engler}, the parameters for the disc density distribution in the model are $R_{\rm 0} = 11.3~$au, $\alpha_{\rm in} = -3.5$, $\alpha_{\rm out} = 6.8$, $z_{0} = 0.6$~au, and $\beta = 0.40$.

For the dust physical properties, we assume a power-law size distribution of grains between $a_{\rm min}$ and $a_{\rm max}$ (i.e. ${\rm d}n \propto a^{-q} {\rm d}a$). We fix the minimum and maximum grain sizes for the dust as $a_{\rm min} = 0.01~\mu$m and $a_{\rm max} = 1~$mm. The exponent of the size distribution $q$ was also fixed to be 3.95. Although a more physically motivated size distribution would omit grains in the range $0.1 < a < 1.0$~$\mu$m from the radiative transfer modelling, due to their rapid removal by radiation pressure, we adopt a single power law size distribution as this is still consistent with the observations as presented in \cite{2012Johnson} and simplifies the modelling process.

For the dust optical properties, we assemble optical constants ($n$,$k$) from the Jena database of optical constants\footnote{\href{https://www.astro.uni-jena.de/Laboratory/Database/databases.html}{Jena Optical Constants Databases}} for the materials used in \cite{2009Lisse} to match the mid-infrared spectrum of HD~172555. Not all the materials utilized in the spectrum fitting presented in table 2 of \cite{2009Lisse} had ($n$,$k$) values available through accessible online databases. The dust composition fit undertaken here should not therefore be directly compared to that work for consistency. Having obtained ($n$,$k$) values for a range of materials, we then extrapolate the optical constants beyond the available wavelength ranges by extrapolation down to 0.1~$\mu$m and up to 2~mm, using the linear slope of the first (or last) two data points for the measured optical constants to extend the range. Using these extrapolated optical constants we run individual disc models through {\sc Hyperion}, assuming the previously determined disc spatial extent and dust size distribution for all materials, an appropriate material density assuming the grains are hard compact spheres, and a scaling density $\rho_{\rm 0}$ of $\approx 10^{-19}~$gcm$^{-3}$ for the individual material SED models which was adjusted so that each individual SED represented emission from the same mass of dust, in this case 3.45$\times10^{23}$~g. 

\begin{figure*}
    \centering
    \includegraphics[width=\textwidth]{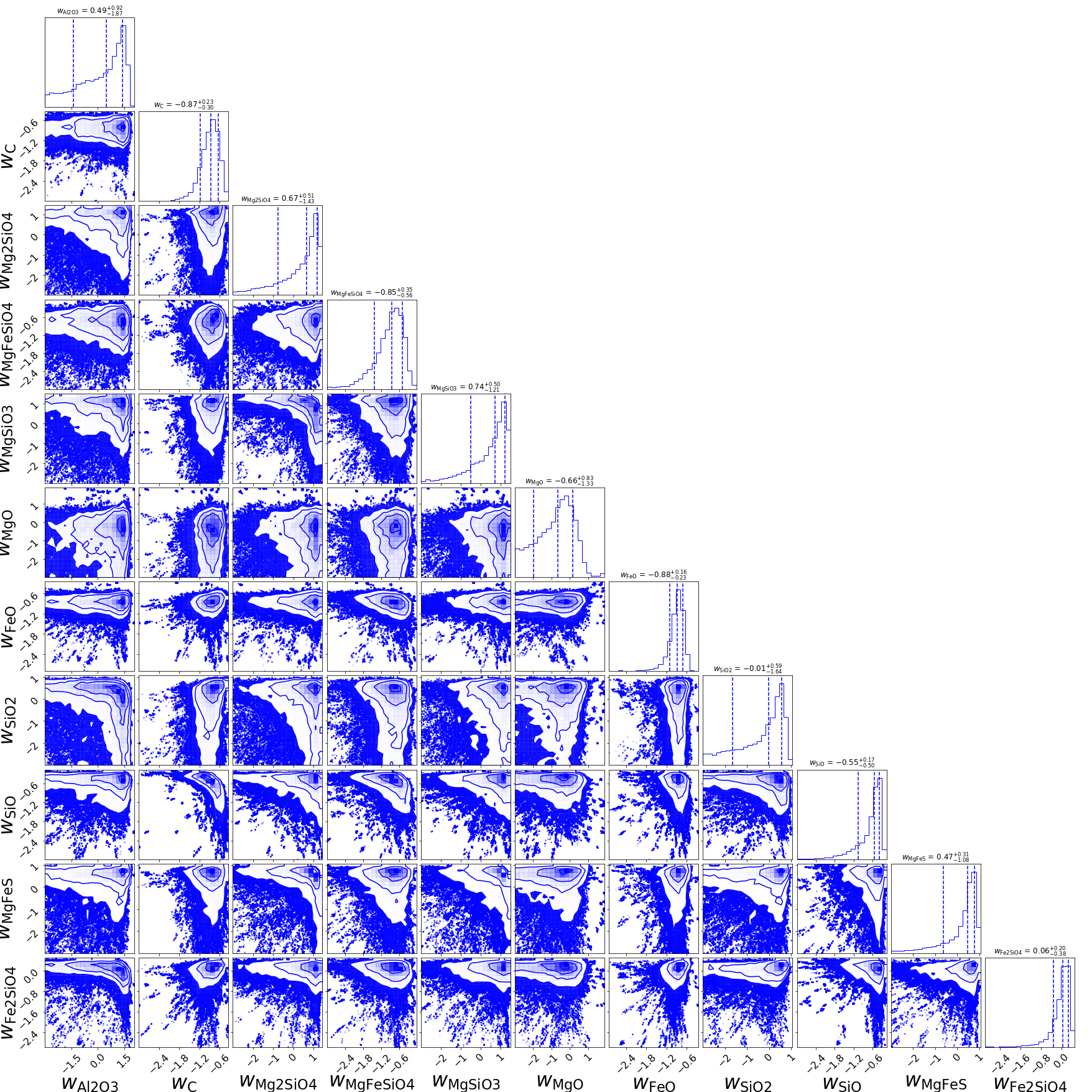}
    \caption{Corner plot showing the probability distributions of the weights of 11 mineral species to HD~172555's SED, see text for details. The species fitted were (shown left to right): Al$_{2}$O$_{3}$, C, Mg$_{2}$SiO$_{4}$, MgFeSiO$_{4}$, MgSiO$_{3}$, MgO, FeO, SiO$_{2}$, SiO, MgFeS, and Fe$_{2}$SiO$_{4}$. The values for the weightings are given in Table \ref{tab:mir_comp}.}
    \label{fig:mcmc}
\end{figure*}

To reproduce the observed SED, we use the affine-invariate Markov Chain Monte Carlo code \textit{emcee} \citep{2013ForemanMackey} to determine the appropriate mixture of the components by assigning each individual SED a weighting factor between 10$^{-3}$ and 10$^{3}$ (i.e. between -3 and 3 in log space) and then calculating the least squares fit of the weighted combination of individual SEDs to the observations. The weighting factors for each species were initialised at 1 (0 in log space), with the individual walkers given a small scatter of 0.05 about this value. We used 200 walkers and 1\,000 steps (200\,000 realisations) to deduce the maximum probability weightings for the materials, discarding the first 400 steps of each chain before calculating the probability distribution to extract the 16$^{\rm th}$, 50$^{\rm th}$, and 84$^{\rm th}$ percentile values. The results of the fitting process are presented in Fig. \ref{fig:mcmc}, with a summary of the weighting factors given in Table \ref{tab:mir_comp}. It should be noted that these weightings are relative to the reference SED of each component, and are therefore not expected to add up to unity. In Appendix \ref{app:composition_seds} we provide SEDs of the individual materials, and a plot of the relative contribution of these components to the total model. 

From the dust composition fitting, we can infer the appropriate optical constants to use in the polarisation spectrum modelling for consistency between all elements of this analysis. The composition analysis is somewhat degenerate, and the relative weightings of different species tied to the sampling of different regions of the mid-infrared spectrum (e.g. denser sampling across the 10~$\mu$m or 20~$\mu$m feature gives different maximum probabilities for the various components). Our fitting results place strong constraints on the weighting on four of the species examined here, namely C, MgFeSiO$_{4}$, SiO, and MgFeS. There are weaker constraints on the weightings for the remaining seven species, several of which have long tails, or broad peaks, to their posterior probability distributions highlighting the degeneracy in the spectral fitting presented here. 

\begin{figure}
    \centering
    \includegraphics[width=0.5\textwidth]{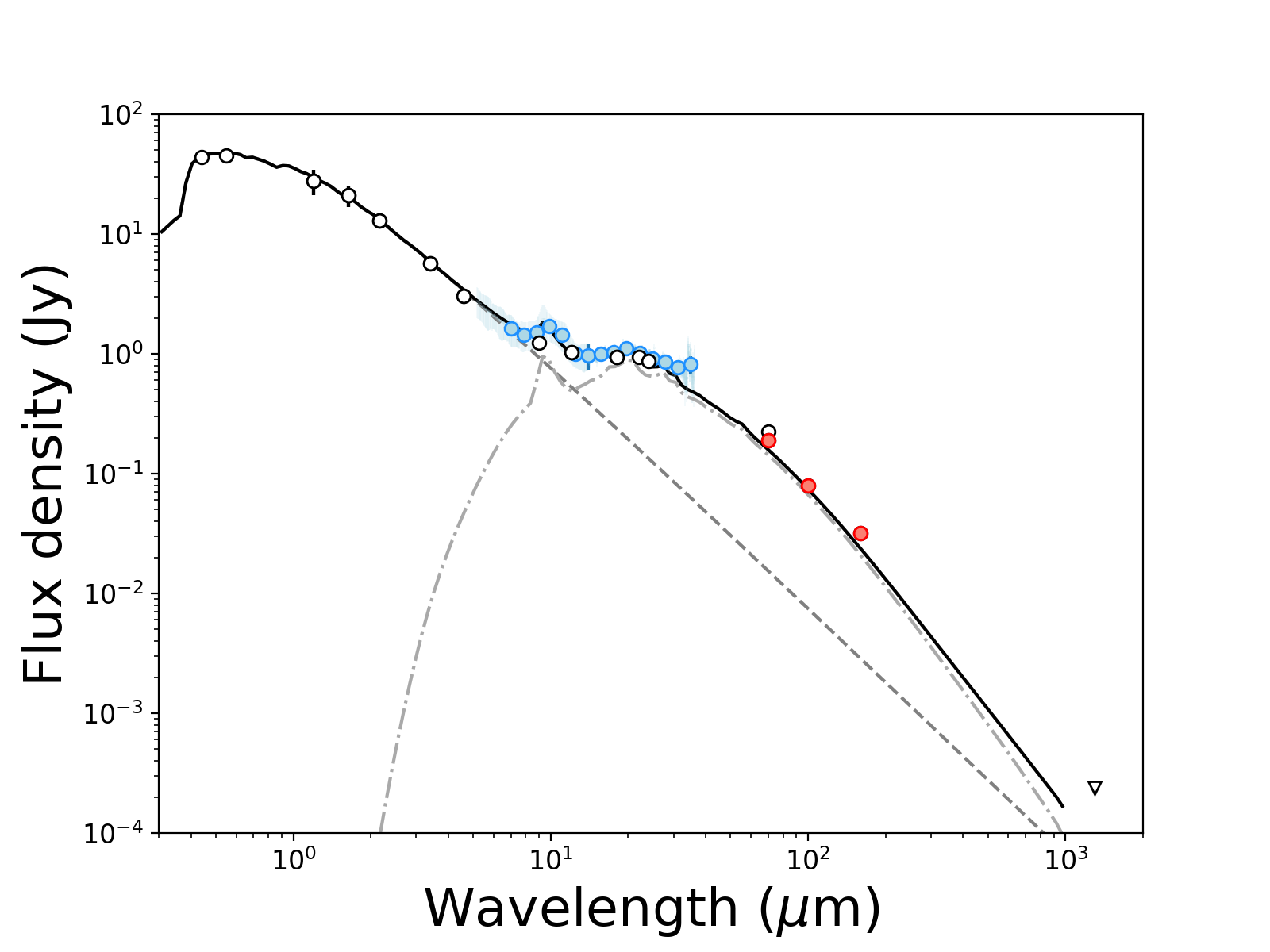}
    \caption{Plot of flux density vs. wavelength for HD~172555. The disc model was calculated based on the disc architecture from the SPHERE images \citep{2018Engler}, and the grain size distribution inferred from fitting the mid-infrared spectrum \citep{2012Johnson}. The stellar photosphere model is represented by the grey dashed line, the total disc contribution (all 11 components) by the grey dot-dash line and the total emission (star+disc) as the solid black line. The blue line represents the \textit{Spitzer}/IRS spectrum \citep{2006Chen}, and photometry extracted from the mid-infrared spectrum are represented as blue circular data points. Red circular data points denote \textit{Herschel}/PACS photometry \citep{2014RiviereMarichalar}.  An upper limit from ALMA Band 6 is denoted by an inverted triangle. \label{fig:sed}}
\end{figure}

A comparison of the model SED with observations of HD~172555 is presented in Fig. \ref{fig:sed}.  The model is consistent, within uncertainties, with the mid-infrared spectrum across its full extent, reproduces the far-infrared photometry, and falls below the millimetre-wavelength upper limit. We have therefore obtained general good agreement between model emission calculated using the weighted sums of the individual dust models and the observations. We have used a single dust grain size distribution with a power law exponent of 3.95 as has been previously found to reproduce the SED for HD~172555's disc \cite[e.g.][]{2006Chen,2009Lisse,2012Johnson}. We find this adequately represents the data in combination with the additional spatial constraint of \cite{2018Engler}.

\begin{table*}
\caption{Photometric measurements of HD~172555 used in the SED fitting. \label{tab:sed}}
\begin{tabular}{lccc}
    \hline\hline
    Wavelength  & Flux Density & Telescope/  & Reference \\
       ($\mu$m) &    (mJy)     & Instrument  &           \\
    \hline
    0.44 & 43915~$\pm$~615 & Johnson $B$ & \cite{2000Hog} \\  
    0.55 & 45112~$\pm$~406 & Johnson $V$ & \cite{2000Hog} \\ 
    1.20  & 28270~$\pm$~7350 & 2MASS $J$ & \cite{2006Skrutskie} \\ 
    1.64  & 21530~$\pm$~4564 & 2MASS $H$ & \cite{2006Skrutskie} \\ 
    2.16 & 12727~$\pm$~394 & 2MASS $K_{S}$ & \cite{2006Skrutskie} \\ 
    3.4  & 5658~$\pm$~1398 & \textit{WISE} W1 & \cite{2010Wright} \\ 
    4.6  & 4038~$\pm$~670 & \textit{WISE} W2 & \cite{2010Wright} \\ 
    9    & 1229~$\pm$~75 & \textit{Akari}/IRC9 & \cite{2010Ishihara} \\ 
    12   & 1032~$\pm$~62 & \textit{WISE} W3 & \cite{2010Wright} \\ 
    18   & 947~$\pm$~52 & \textit{Akari}/IRC18 & \cite{2010Ishihara} \\ 
    22   & 1520~$\pm$~525 & \textit{WISE} W4 & \cite{2010Wright} \\ 
    24   & 947~$\pm$~83 & \textit{Spitzer}/MIPS24 & \cite{2014Chen} \\ 
    70   & 226~$\pm$~16 & \textit{Spitzer}/MIPS70 & \cite{2012Johnson} \\
    70   & 191~$\pm$~13 & \textit{Herschel}/PACS70 & \cite{2014RiviereMarichalar} \\
    100  & 79~$\pm$~6 & \textit{Herschel}/PACS100 & \cite{2014RiviereMarichalar} \\ 
    160  & 32~$\pm$~2 & \textit{Herschel}/PACS160 & \cite{2014RiviereMarichalar} \\ 
    1300 & $<$0.240      & ALMA Band 6      & This work \\
    \hline
\end{tabular}
\end{table*}

\begin{table}
    \centering
    \caption{Summary of the material weights determined by fitting of a composite SED of 11 materials to the SED and mid-infrared spectrum. Total dust mass of the disc (in grains up to 1~mm) is inferred to be 1.10$\times10^{-3}~M_{\oplus}$, or 7.19$\times10^{21}$~kg. \label{tab:mir_comp}}
    \begin{tabular}{l|l|l}
    \hline
    Species     &  $\log$(Wgt) & Mass ($10^{-5}M_{\oplus}$) \\
    \hline\hline
    Aluminium Oxide, Al$_{2}$O$_{3}$  & \phantom{-}0.49$^{+0.93}_{-1.90}$ & 17.83$^{+133.15}_{-17.61}$ \\
    Carbon, C & -0.88$^{+0.24}_{-0.31}$ & 0.77$^{+0.55}_{-0.39}$\\
    Forsterite, Mg$_{2}$SiO$_{4}$  & \phantom{-}0.66$^{+0.51}_{-1.46}$ & 26.70$^{+60.60}_{-25.77}$\\
    Olivine, MgFeSiO$_{4}$  & -0.85$^{+0.36}_{-0.57}$ & 0.81$^{+1.04}_{-0.59}$\\
    Fayalite, Fe$_{2}$SiO$_{4}$  & \phantom{-}0.05$^{+0.20}_{-0.38}$ & 6.51$^{+3.82}_{-3.80}$\\
    Ortho-Enstatite, MgSiO$_{3}$  & \phantom{-}0.73$^{+0.51}_{-1.22}$ & 31.04$^{+69.25}_{-29.19}$\\
    Magnesium Oxide, MgO & -0.68$^{+0.87}_{-1.34}$ & 1.19$^{+7.09}_{-1.13}$\\
    Iron Oxide, FeO  & -0.89$^{+0.16}_{-0.23}$ & 0.75$^{+0.34}_{-0.31}$\\
    Silicon Dioxide, SiO$_{2}$ & -0.02$^{+0.59}_{-1.66}$ & 5.54$^{+16.04}_{-5.24}$\\
    Silicon Oxide, SiO & -0.55$^{+0.17}_{-0.50}$ & 1.63$^{+0.78}_{-1.12}$\\
    Magnesium Iron Sulphide, MgFeS  & \phantom{-}0.47$^{+0.30}_{-1.10}$ & 16.98$^{+17.24}_{-15.62}$\\
    \hline
    \end{tabular}
\end{table}

\subsection{Polarimetry}

We model the HIPPI/-2 aperture polarisation measurements using the disc architecture and orientation of the scattered light disc imaged in \cite{2018Engler}. The albedo and scattering M\"uller matrix for each composition was calculated using a parallelized Fortran95 implementation of BHMie\footnote{\href{https://github.com/pscicluna/bhmie}{https://github.com/pscicluna/bhmie}}. We calculate the disc polarisation using a 3D model of the disc density distribution combined with the dust optical constants, following the model presented in \cite{2015Scicluna}. We assume the measured polarisation is produced in single scattering events for dust grains in the optically thin disc redirecting unpolarised starlight toward the observer. The scattering angles for a 3D volume containing the disc (centred on the star) are calculated to generate linear fractional polarisation ($q$ and $u$) values at 200 wavelengths between 0.3 and 0.9~$\mu$m. Polarisation measurements of the disc (as projected onto the plane of the sky) were determined using the ratio of M\"uller matrix elements $S_{11}$ and $S_{12}$, the dust albedo, and the stellar flux. The orientation of the disc in the model is chosen such that the degree of polarisation in the U vector is zero by default. 

We proceed by generating model polarisation spectra for each dust composition spanning 0.3 to 0.9~$\mu$m, sampling 200 wavelengths within this range, for minimum dust grains sizes between 0.01 and 6.5~$\mu$m. We fixed the maximum grain size as 1~mm, with a power law size distribution exponent of -3.95, as per the radiative transfer modelling. We did not omit grains with sizes 0.1 to 1.0~$\mu$m from the scattering polarisation calculation despite these grains being swiftly removed by radiation forces. If such grains were shown to be capable of reproducing the observed polarisation it would be an intriguing result given that they would be expected to be transitory in the disc around HD~172555.

Whilst the radiative transfer modelling used grains smaller than $0.3~\mu$m to generate the dust emission, such grains are weakly polarising \citep{2016Marshall,2017Kirchschlager}, and would not be expected to produce an adequate fit to the polarimetry, so we curtail the parameter space of our search accordingly. Similarly, dust grains larger than 5~$\mu$m would not be expected to produce strongly polarised light at shorter wavelengths setting the upper boundary to the range of grain sizes considered here. We also calculate polarisation spectra for astronomical silicate dust grains \citep{2003Draine} as a point of comparison to the dust composition determined from this work.

We calculate a single model polarisation spectrum for each grain size using the weightings calculated in the dust grain radiative transfer modelling. The best-fit grain size was then determined using a least-squares fit of the model polarisation spectrum for each grain size to the multi-wavelength polarimetry. A bandpass model was used to calculate synthetic HIPPI/-2 polarimetric measurements across the model spectra at the relevant wavelengths for the fitting process.

\begin{figure*}
    \centering
    \includegraphics[width=0.48\textwidth]{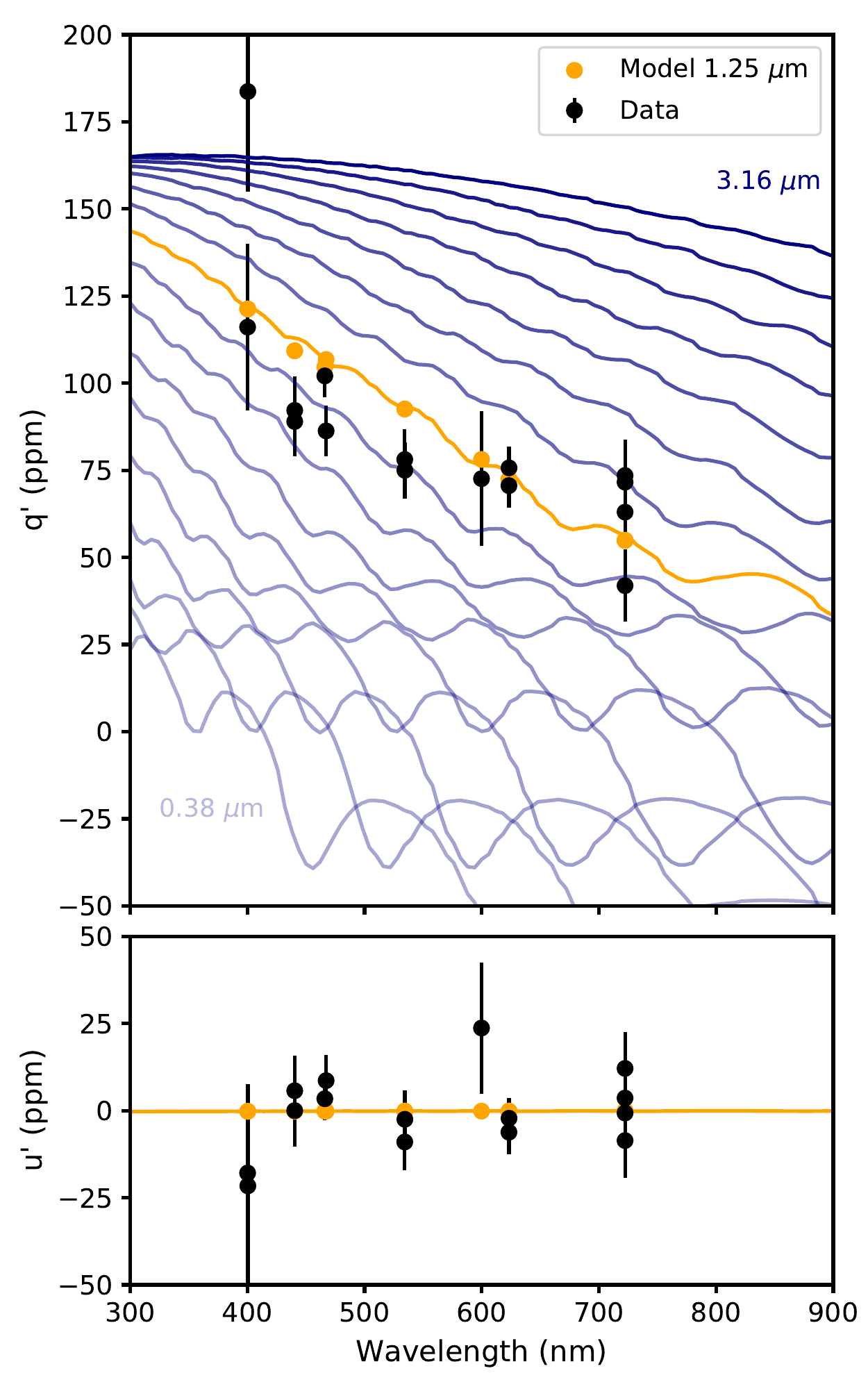}
    \includegraphics[width=0.48\textwidth]{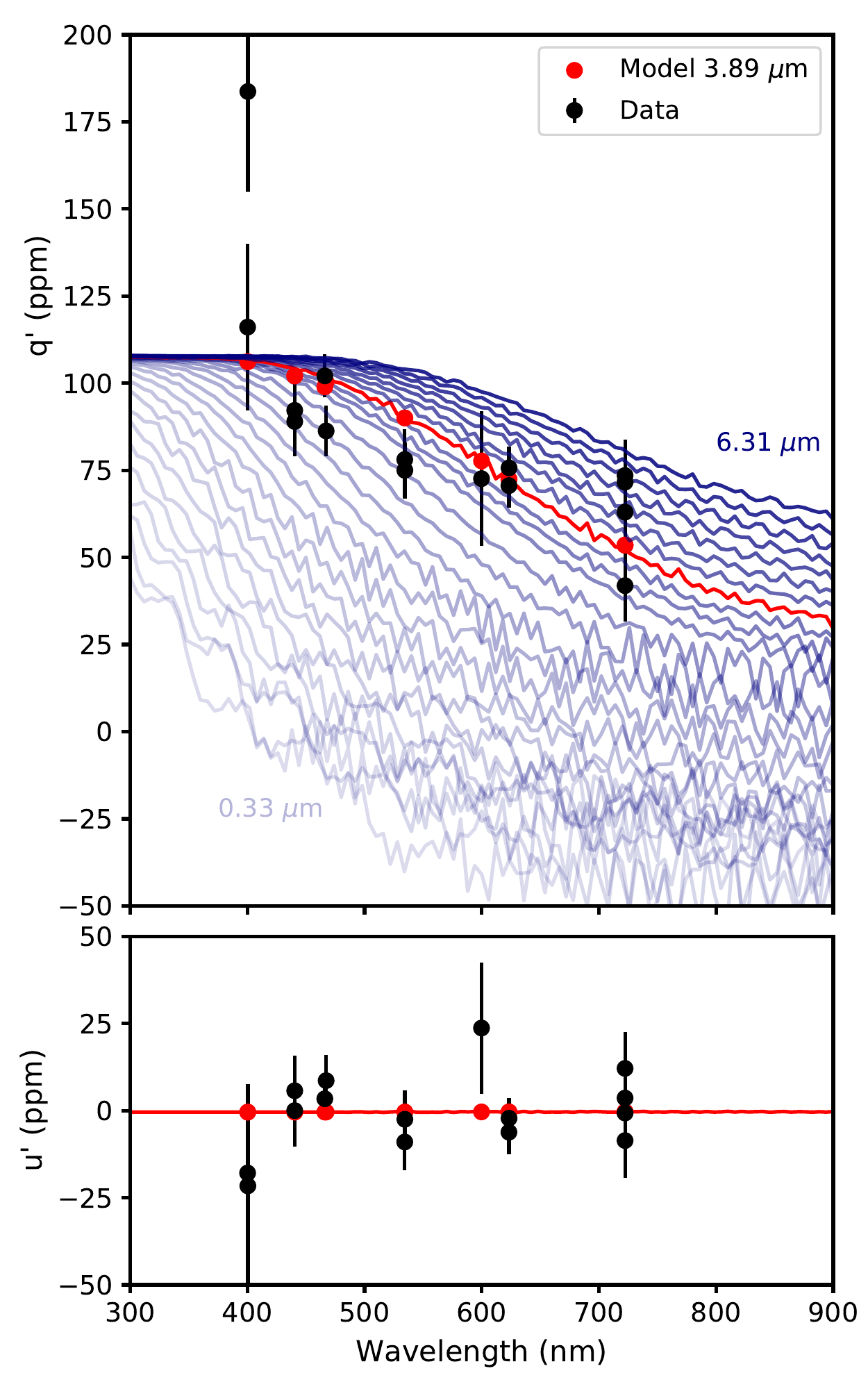}
    \caption{Plots of wavelength vs. linear polarisation vectors $q^{\prime}$ (top) and $u^{\prime}$ (bottom) for the best fit models of astronomical silicate (left) and radiative transfer derived dust composition (right) following the assumed disc architecture and dust grain properties (see text for details). The polarisation spectra for the best-fitting dust grain size for astronomical silicate (mixed composition) are denoted by the yellow (red) line, with the waveband derived polarimetric measurements denoted by coloured dots. Polarisation spectra for adjacent dust grain sizes are given as shaded blue lines ([0.33,] 0.38, 0.43, 0.49, 0.56, 0.64, 0.73, 0.84, 0.96, 1.09, 1.25, 1.43, 1.63, 1.86, 2.12, 2.42, 2.77, 3.16, [3.39, 3.63, 3.89, 4.17, 4.47, 4.79, 5.13, 5.50, 5.89, 6.31] $\mu m$), and the observations are shown as black dots.}
    \label{fig:pol}
\end{figure*}

We found a satisfactory fit to the polarisation using dust grains with $n$,$k$ values calculated from a dust composition determined from the relative weights of the 11 components used in the SED fitting. The best-fit polarisation model using this dust composition has a minimum grain size $a_{\rm min} = 3.89~\mu$m and is an adequate match to the observations, as shown in Fig. \ref{fig:pol}. The model polarisation spectra do not replicate the steep rise in polarimetric signal toward shorter wavelengths, but are consistent within the substantial uncertainties attached to the magnitude and shape of the model polarisation spectra at those shortest wavelengths. 

With the astronomical silicate models, we find a best-fit minimum grain size $a_{\rm min} = 1.25~\mu$m from a least squares fit between the grain size dependent polarisation spectra and the observations. The dust grain size we infer from this simple model is comparable to the radiation blow-out grain size for HD~172555 ($a_{\rm blow} \approx 1.5~\mu$m) and consistent with the best-fit grain size derived from the radiative transfer models. From this result, combined with that of the mixed composition, we infer that the dust grains responsible for the polarisation are not the very smallest grains which dominate the overall disc surface area and from which the mid-infrared spectral features originate.

\section{Discussion}
\label{sec:dis}

\subsection{Technique:}

Previously published aperture polarisation measurements at the $10^{-5}$ to $10^{-4}$ r.m.s. polarisation level have not convincingly detected signatures of dust polarisation from debris disc host stars \citep{2015Garcia, 2019Vandenportal}. The fundamental difficulty in applying the technique to debris discs is separating the interstellar polarisation from that due to the debris disc. Ordinarily the fractional polarisation from a debris disc system will not exceed half of the associated continuum flux excess. A dust fractional excess ($L_{\rm dust}/L_{\star}$) of 10$^{-4}$ would only be expected to yield a polarimetric signal of a few $\times10^{-5}$ (depending on the assumed grain shape and porosity). This means that only the brightest discs have been detectable up until now. Scattered light detections of debris discs have established that the continuum emission and scattered light from a given disc do not always correlate \citep[e.g.][]{2014Schneider}, and debris dust grains generally have low albedoes ($< 0.1$) comparable to that of bodies in the Edgeworth-Kuiper belt \citep{2018Choquet,2018Marshall}. These factors reduce the expected scattered light brightness of debris dust, further reducing their overall detectability in polarisation.

The development of parts-per-million ($10^{-6}$) polarimeters has helped with this \citep{2010Bailey, 2010aWiktorowicz, 2010bWiktorowicz, 2016Cotton, 2017bCotton, 2020bCotton}, but interstellar polarisation contributions are typically tens of ppm within the Local Hot Bubble (within 75 to 150~pc, \citealp{2010Bailey, 2016Cotton, 2017bCotton, 2020Piirola}), and hundreds of ppm just beyond it \citep{2019aCotton, 2019Gontcharov}. Consequently, to characterise the disc polarisation, the interstellar polarisation must be removed. The task is made more difficult since the wavelength dependence of interstellar polarisation within the Local Hot Bubble is not yet well defined \citep{2016Marshall, 2019aCotton}. Our results with regard to the interstellar medium here, where we obtain $\lambda_{\rm max}=472.6 \pm 79.0$~nm for HD~162521 and HD~167425 combined, serve to improve this situation. 

A similar problem exists in aperture polarimetry of other non-variable phenomena, for instance in determining the inclination and rotation rate of rapidly rotating stars \citep{2017aCotton,2020bBailey}. That case is more easily dealt with however, since the wavelength dependence of polarisation intrinsic to the system is well defined \citep{2020bBailey}. Whereas in debris disc systems we do not yet understand their architectures or compositions well enough to so easily constrain the parameters; consequently, observations of the target system alone are not sufficient.

Imaging polarimetry of discs has been possible for the brightest discs with favourable (generally edge-on) orientations to the line-of-sight for some time, e.g. $\beta$~Pic \citep{1991Gledhill, 2000Krivova, 2006Tamura} and AU~Mic \citep{2007Graham}. New instruments such as the Gemini Planet Imager (e.g., \citealp{2020Esposito}) and SPHERE-ZIMPOL (e.g. HR~4796A, \citealp{2019Milli}) have increased the number of such systems investigated. As related here for HD~172555, these allow the removal of the interstellar component. Furthermore, by occluding the largely unpolarised central star they reduce its fractional contribution to insignificance. Of course being able to map polarisation through the disc has its own advantages. However, there are also disadvantages; imaging is easier at red and infrared wavelengths where adaptive optics can be effective; and the occulting disc will obscure the dust nearest the star along with it. Multi-band measurements at optical wavelengths are particularly important for elucidating the nature of the dust in disc systems. 

\subsection{Dust properties:}

We have attempted to match the multi-wavelength aperture polarimetry measurements reported here using a simple polarisation model with dust grain scattering properties based on the size distribution and optical properties determined by Mie theory based on both the composition inferred from the radiative transfer modelling and astronomical silicate.

Our model manages to replicate the overall wavelength dependence of the observed polarisation properties of the disc except for the shortest wavelengths. We have thus been able to find a single comprehensive solution for the dust continuum emission and polarisation observations for HD~172555. This is perhaps surprising given the number of simplifying assumptions, including the adoption of Mie theory to calculate the dust grain optical properties which is well known to poorly replicate polarimetric dust signatures. The best-fit minimum grain size with a dust composition based on the radiative transfer modelling was $a_{\rm min} = 3.89~\mu$m. Using the astronomical silicate composition, we obtain a best-fit minimum grain size of 1.25~$\mu$m.  Our analysis is strongly suggestive that $\sim$ micron-sized grains, larger than the radiation blow-out size, are responsible for the observed polarisation. This rules out small, sub-micron grains which are responsible for the mid-infrared spectral features as the origin of the polarisation. This is consistent with the generally seen result in SED modelling that the minimum grain size around later type stars is a few times the blow out limit from radiation pressure. 

However, the assumptions inherent in this work used to calculate the optical constants from the composition model and regarding the dust grain shape prevent us placing a tight constraint on the exact size of dust grains responsible for the polarimetric signal. In particular, $n$,$k$ measurements for dust species spanning the full range of wavelengths present in the observations were not always available. This required extrapolation to fill in the blanks, in some cases across important regions for interpreting scattered light and polarisation. Comprehensive, broad wavelength (0.1~$\mu$m to 1~mm) coverage of dust optical constants for astronomically interesting species would be an invaluable resource, particularly across a range of temperatures relevant for circumstellar disc studies (10s to 100s K).

\section{Conclusions}
\label{sec:con}

We have undertaken a multi-wavelength study of the dust continuum emission and optical polarimetric properties of the HD~172555 debris disc system. We determine that the previously posited dust composition and grain size distribution \citep{2006Chen,2009Lisse,2012Johnson} are consistent with the new spatial constraint imposed by scattered light imaging \citep{2018Engler}. 

We have convincingly detected polarisation from a debris disc using aperture polarimetry for the first time, demonstrating the consistency of our measurements with those of imaging polarisation at comparable wavelengths. The measurements presented here illustrate the need for parts-per-million sensitivity to undertake such an analysis. 

We constrain the dust grain size responsible for the observed polarisation (under the assumption of Mie theory and astronomical silicate composition) to be around 1.25~$\mu$m, and 3.89~$\mu$m for grains with optical constants derived from the dust composition used to fit the spectral energy distribution. The smallest grains in the disc, i.e. those responsible for the mid-infrared spectral features and which dominate the disc surface area, are not the ones responsible for the observed polarisation from the disc. 
We are able to replicate the polarimetric observations of HD~172555 with our model using a single dust composition and grain size with a composition derived from the radiative transfer modelling. However, this result requires more rigorous testing with scattering from non-spherical dust grains and a more realistic grain size distribution including the effects of radiation pressure, for example. Here we have demonstrated a unique methodology that uses the complimentary nature of imaging and aperture techniques to refine our understanding of the debris dust in this system. As such this is a showcase of the synergy of imaging and aperture polarimetric techniques.

\section*{Acknowledgements}

This is a pre-copyedited, author-produced PDF of an article accepted for publication in MNRAS  following peer review. The version of record Marshall et al., 2020, MNRAS, 499, 4, 5915--5931 is available online \href{https://academic.oup.com/mnras/article/499/4/5915/5925356}{here}.

We thank the anonymous referee for their constructive criticism which helped improve the manuscript.

This research has made use of the SIMBAD database, operated at CDS, Strasbourg, France. 

This research has made use of NASA's Astrophysics Data System.

This paper has made use of the Python packages {\sc astropy} \citep{2013AstroPy,2018AstroPy}, {\sc SciPy} \citep{SciPy}, {\sc matplotlib} \citep{2007Hunter}, \textit{emcee} \citep{2013ForemanMackey}, {\sc corner} \citep{2016ForemanMackey},and {\sc Hyperion} \citep{2011Robitaille}.

We thank Dag Evensberget for assisting with observations in August of 2018.

The development of HIPPI was funded by the Australian Research Council through Discovery Projects grant DP140100121 and by the UNSW Faculty of Science through its Faculty Research Grants program. Funding for the construction of HIPPI-2 was provided by UNSW through the Science Faculty Research Grants Program. We thank the staff of the AAT for their support for our observing programme. 

JPM acknowledges research support by the Ministry of Science and Technology of Taiwan under grants MOST104-2628-M-001-004-MY3, MOST107-2119-M-001-031-MY3, and MOST109-2112-M-001-036-MY3, and Academia Sinica under grant AS-IA-106-M03.

This paper makes use of the following ALMA data: ADS/JAO.ALMA\#2013.1.01147.S. ALMA is a partnership of ESO (representing its member states), NSF (USA) and NINS (Japan), together with NRC (Canada), MOST and ASIAA (Taiwan), and KASI (Republic of Korea), in cooperation with the Republic of Chile. The Joint ALMA Observatory is operated by ESO, AUI/NRAO and NAOJ. The ALMA data used in this work were retrieved from the JVO portal (http://jvo.nao.ac.jp/portal/) operated by ADC/NAOJ.

\section*{Data availability}

The data underlying this article are available in the article and in its online supplementary material.




\bibliographystyle{mnras}
\bibliography{refs} 


\appendix

\section{Standard Observations}
\label{sec:std_obs}

The instrumental PA is calibrated with reference to high polarisation standard stars observed in \gfil~or without a filter. Table \ref{tab:pa_prec} lists the standards observed for each run. The errors associated with the literature values are around a degree. The column labelled `S.D.' gives the standard deviation of $\Delta PA=PA_{obs}-PA_{lit}$, where $PA_{obs}$ is the $\theta$ for the observation after calibration, and $PA_{lit}$ the literature value as given in \citet{2020Bailey}. Table \ref{tab:pa_multi} shows that for the 2018AUG run 425SP and 500SP observations, after calibration based on \gfil/Clear, are significantly rotated. A second order correction was therefore applied.

\begin{table}
\caption{Precision in PA by Observing Run.}
\centering
\tabcolsep 4 pt
\begin{tabular}{lc|ccccc|r}
\hline
Run &  \multicolumn{5}{c|}{PA Standard Observations} & S.D.\\
\multicolumn{1}{r}{HD:} & 23512 & 147084 & 154445 & 160529 & 187929 & \multicolumn{1}{|c}{($^{\circ}$)} \\
\hline
2015MAY & 0 & 4 & 1 & 0 & 0 & 0.17 \\
2015OCT & 1 & 0 & \textit{2} & 0 & \textit{2} & 0.24 \\
2017JUN & 0 & 2 & 1 & 1 & 0 & 1.11 \\
2018AUG & 0 & 3 & 0 & 3 & \textit{5} & 0.86 \\
\hline
\end{tabular}
\begin{flushleft}
Note: All standards were observed in \gfil~except where the number is italicised, in which case one instance was observed without a filter.
\end{flushleft}
\label{tab:pa_prec}
\end{table}

\begin{table}
\caption{Multi-band PA Correction for 2018AUG Run.}
\centering
\tabcolsep 3 pt
\begin{tabular}{lllc|ccc|rr}
\hline
\multicolumn{2}{c}{Band}& $\lambda_{\rm eff}$ &  \multicolumn{4}{c|}{PA Standard Observations} & $\Delta$PA & S.D.\\
&& (nm) & \multicolumn{1}{r}{HD:} & 147084 & 160529 & 187929 & \multicolumn{1}{|c}{($^{\circ}$)} & \multicolumn{1}{|c}{($^{\circ}$)}\\
\hline
425SP   & B & 403.5 && 1 & 2 & 3 &  2.60 & 1.08 \\
500SP   & B & 441.3 && 1 & 2 & 2 &  5.80 & 1.27 \\
\gfil   & B & 466.9 && 2 & 2 & 3 & -0.09 & 0.87 \\
\gfil   & R & 485.7 && 1 & 1 & 1 & -0.03 & 1.21 \\
V       & B & 537.6 && 0 & 2 & 2 & -0.06 & 0.84 \\
\rfil   & B & 603.7 && 1 & 2 & 2 &  0.64 & 0.69 \\
\rfil   & R & 611.2 && 1 & 1 & 1 &  0.10 & 0.85 \\
650LP   & R & 722.4 && 1 & 1 & 1 &  0.30 & 0.70 \\
\hline
\end{tabular}
\begin{flushleft}
Note: The observations cover a number of different modulator performance eras.\\ 
\end{flushleft}
\label{tab:pa_multi}
\end{table}

Tables \ref{tab:lp_std} and \ref{tab:lp_std2} list all observations used to calibrate the TP for each run in each band. The polarisation of each of these stars is assumed to be zero.

\begin{table*}
\caption{HIPPI low polarisation standard observations. \label{tab:lp_std}}
\tabcolsep 5 pt
\begin{tabular}{lcccrcccccrr}
    \hline\hline
    Standard    & UT & Run & Instr. & Ap. & Mod. & Fil & PMT & $\lambda_{\rm eff}$ & Eff & \multicolumn{1}{c}{q} & \multicolumn{1}{c}{u} \\
              & & & & (\arcsec) & Era & & & (nm) && \multicolumn{1}{c}{(ppm)} & \multicolumn{1}{c}{(ppm)}    \\
    \hline
HD  48915	&	2015-05-24 09:09:47	&	2015MAY	&	HIPPI	&	6.6	&	E1	&	425SP	&	B	&	400.6	&	0.570	&	-67.9	$\pm$	\04.1	&	3.9	$\pm$	\03.9	\\
HD 140573	&	2015-05-23 13:16:51	&	2015MAY	&	HIPPI	&	6.6	&	E1	&	425SP	&	B	&	407.8	&	0.602	&	-52.0	$\pm$	17.7	&	-1.1	$\pm$	17.4	\\
HD 140573	&	2015-05-25 13:43:21	&	2015MAY	&	HIPPI	&	6.6	&	E1	&	425SP	&	B	&	407.8	&	0.602	&	-28.7	$\pm$	17.8	&	18.5	$\pm$	17.4	\\
HD 140573	&	2015-05-26 13:07:50	&	2015MAY	&	HIPPI	&	6.6	&	E1	&	425SP	&	B	&	407.8	&	0.602	&	-45.2	$\pm$	14.7	&	10.2	$\pm$	14.9	\\
HD 140573	&	2015-06-26 11:43:37	&	2015JUN	&	HIPPI	&	6.6	&	E1	&	425SP	&	B	&	407.8	&	0.602	&	-24.4	$\pm$	15.7	&	13.7	$\pm$	15.9	\\
HD 140573	&	2015-06-27 13:09:02	&	2015JUN	&	HIPPI	&	6.6	&	E1	&	425SP	&	B	&	408.1	&	0.603	&	-38.8	$\pm$	14.8	&	66.1	$\pm$	14.9	\\
\multicolumn{2}{l}{\textit{Adopted TP}}		&		&		&		&		&		&		&	406.6	&		&	-42.8	$\pm$	\06.1	&	18.5	$\pm$	\06.0	\\
\hline																											
HD  48915	&	2015-05-23 07:58:16	&	2015MAY	&	HIPPI	&	6.6	&	E1	&	\gfil	&	B	&	465.6	&	0.895	&	-38.7	$\pm$	\01.3	&	-2.1	$\pm$	\01.3	\\
HD  48915	&	2015-05-24 07:58:30	&	2015MAY	&	HIPPI	&	6.6	&	E1	&	\gfil	&	B	&	465.8	&	0.896	&	-39.8	$\pm$	\00.7	&	-0.1	$\pm$	\00.7	\\
HD 102647	&	2015-06-27 08:28:37	&	2015JUN	&	HIPPI	&	6.6	&	E1	&	\gfil	&	B	&	466.5	&	0.897	&	-44.0	$\pm$	\02.3	&	1.0	$\pm$	\02.3	\\
HD 140573	&	2015-05-22 11:55:52	&	2015MAY	&	HIPPI	&	6.6	&	E1	&	\gfil	&	B	&	476.3	&	0.918	&	-23.8	$\pm$	\04.9	&	18.9	$\pm$	\05.1	\\
HD 140573	&	2015-05-26 12:30:43	&	2015MAY	&	HIPPI	&	6.6	&	E1	&	\gfil	&	B	&	475.9	&	0.917	&	-36.0	$\pm$	\03.9	&	6.9	$\pm$	\03.9	\\
HD 140573	&	2015-06-26 12:11:26	&	2015JUN	&	HIPPI	&	6.6	&	E1	&	\gfil	&	B	&	475.9	&	0.917	&	-41.7	$\pm$	\04.2	&	-9.1	$\pm$	\04.4	\\
HD 140573	&	2015-06-27 12:45:15	&	2015JUN	&	HIPPI	&	6.6	&	E1	&	\gfil	&	B	&	475.9	&	0.917	&	-27.2	$\pm$	\03.8	&	18.7	$\pm$	\03.8	\\
\multicolumn{2}{l}{\textit{Adopted TP}}			&		&		&		&		&		&		&	471.7	&		&	-35.9	$\pm$	\01.3	&	4.9	$\pm$	\01.3	\\
\hline																											
HD  48915	&	2015-05-24 08:14:17	&	2015MAY	&	HIPPI	&	6.6	&	E1	&	\rfil	&	B	&	598.9	&	0.835	&	-25.0	$\pm$	\01.6	&	2.5	$\pm$	\01.7	\\
HD 140573	&	2015-05-23 13:42:01	&	2015MAY	&	HIPPI	&	6.6	&	E1	&	\rfil	&	B	&	603.7	&	0.826	&	-31.1	$\pm$	\07.7	&	-5.0	$\pm$	\08.0	\\
HD 140573	&	2015-05-25 13:20:36	&	2015MAY	&	HIPPI	&	6.6	&	E1	&	\rfil	&	B	&	603.7	&	0.826	&	-34.0	$\pm$	\06.7	&	0.2	$\pm$	\06.7	\\
\multicolumn{2}{l}{\textit{Adopted TP}}		&		&		&		&		&		&		&	602.1	&		&	-30.0	$\pm$	\03.4	&	-0.7	$\pm$	\03.5	\\
\hline																											
HD   2151	&	2015-10-14 09:42:14	&	2015OCT	&	HIPPI	&	6.6	&	E1	&	\gfil	&	B	&	472.5	&	0.910	&	-56.0	$\pm$	\03.8	&	-2.2	$\pm$	\03.8	\\
HD   2151	&	2015-10-19 13:07:41	&	2015OCT	&	HIPPI	&	6.6	&	E1	&	\gfil	&	B	&	472.0	&	0.909	&	-49.8	$\pm$	\03.7	&	-3.9	$\pm$	\03.7	\\
HD   2151	&	2015-10-29 09:36:13	&	2015NOV	&	HIPPI	&	6.6	&	E1	&	\gfil	&	B	&	472.3	&	0.909	&	-54.2	$\pm$	\03.8	&	1.9	$\pm$	\03.8	\\
HD  48915	&	2015-10-16 18:27:29	&	2015OCT	&	HIPPI	&	6.6	&	E1	&	\gfil	&	B	&	464.7	&	0.893	&	-51.6	$\pm$	\00.7	&	1.9	$\pm$	\00.7	\\
HD  48915	&	2015-10-19 17:16:09	&	2015OCT	&	HIPPI	&	6.6	&	E1	&	\gfil	&	B	&	464.9	&	0.894	&	-48.9	$\pm$	\01.0	&	-2.9	$\pm$	\01.5	\\
HD  48915	&	2015-11-02 18:08:07	&	2015NOV	&	HIPPI	&	6.6	&	E1	&	\gfil	&	B	&	464.7	&	0.893	&	-42.0	$\pm$	\00.9	&	4.4	$\pm$	\00.9	\\
\multicolumn{2}{l}{\textit{Adopted TP}}		&		&		&		&		&		&		&	468.5	&		&	-50.4	$\pm$	\01.1	&	-0.1	$\pm$	\01.1	\\
\hline																									
HD   2151	&	2017-06-25 19:36:07	&	2017JUN	&	HIPPI	&	6.6	&	E2	&	\gfil	&	B	&	472.0	&	0.888	&	-21.3	$\pm$	\04.2	&	6.8	$\pm$	\04.1	\\
HD   2151	&	2017-08-10 19:05:57	&	2017AUG	&	HIPPI	&	6.6	&	E2	&	\gfil	&	B	&	472.3	&	0.888	&	-16.9	$\pm$	\04.2	&	4.4	$\pm$	\04.6	\\
HD  48915	&	2017-08-11 19:40:30	&	2017AUG	&	HIPPI	&	6.6	&	E2	&	\gfil	&	B	&	466.2	&	0.872	&	-21.4	$\pm$	\04.8	&	-10.3	$\pm$	\05.0	\\
HD  48915	&	2017-08-19 19:00:40	&	2017AUG	&	HIPPI	&	6.6	&	E2	&	\gfil	&	B	&	466.2	&	0.872	&	2.6	$\pm$	\02.7	&	-5.7	$\pm$	\02.6	\\
HD 102647	&	2017-06-22 09:03:49	&	2017JUN	&	HIPPI	&	6.6	&	E2	&	\gfil	&	B	&	466.5	&	0.873	&	-3.1	$\pm$	\02.4	&	0.7	$\pm$	\02.6	\\
HD 102647	&	2017-06-30 08:26:33	&	2017JUN	&	HIPPI	&	6.6	&	E2	&	\gfil	&	B	&	466.5	&	0.873	&	-4.7	$\pm$	\02.5	&	-19.9	$\pm$	\02.5	\\
HD 102870	&	2017-06-23 08:58:12	&	2017JUN	&	HIPPI	&	6.6	&	E2	&	\gfil	&	B	&	471.5	&	0.886	&	-10.9	$\pm$	\05.2	&	15.5	$\pm$	\04.9	\\
HD 102870	&	2017-06-25 08:22:44	&	2017JUN	&	HIPPI	&	6.6	&	E2	&	\gfil	&	B	&	471.3	&	0.886	&	-3.1	$\pm$	\05.9	&	-10.5	$\pm$	\05.4	\\
\multicolumn{2}{l}{\textit{Adopted TP}}		&		&		&		&		&		&		&	469.1	&		&	-9.9	$\pm$	\01.5	&	-2.4	$\pm$	\01.5	\\
\hline																										    \hline
\end{tabular}
\end{table*}

\begin{table*}
\caption{HIPPI-2 low polarisation standard observations. \label{tab:lp_std2}}
\tabcolsep 5 pt
\begin{tabular}{lcccrcccccrr}
    \hline\hline
    Standard    & UT & Run & Instr. & Ap. & Mod. & Fil & PMT & $\lambda_{\rm eff}$ & Eff & \multicolumn{1}{c}{q} & \multicolumn{1}{c}{u} \\
              & & & & (\arcsec) & Era & & & (nm) && \multicolumn{1}{c}{(ppm)} & \multicolumn{1}{c}{(ppm)}    \\
    \hline	
HD   2151	&	2018-07-11 19:00:22	&	2018JUL	&	HIPPI-2	&	11.9	&	E4	&	500SP	&	B	&	446.4	&	0.717	&	-7.7	$\pm$	3.8	&	32.8	$\pm$	4.2	\\
HD   2151	&	2018-09-02 16:38:53	&	2018AUG	&	HIPPI-2	&	11.9	&	E7	&	500SP	&	B	&	446.4	&	0.474	&	35.7	$\pm$	6.9	&	19.0	$\pm$	7.0	\\
HD  10700	&	2018-07-10 18:51:19	&	2018JUL	&	HIPPI-2	&	11.9	&	E4	&	500SP	&	B	&	448.2	&	0.735	&	-9.5	$\pm$	5.4	&	7.1	$\pm$	5.1	\\
HD  10700	&	2018-08-28 16:09:19	&	2018AUG	&	HIPPI-2	&	11.9	&	E7	&	500SP	&	B	&	448.2	&	0.490	&	34.7	$\pm$	7.5	&	36.3	$\pm$	8.1	\\
HD  48915	&	2018-08-19 19:33:02	&	2018AUG	&	HIPPI-2	&	11.9	&	E5	&	500SP	&	B	&	438.6	&	0.547	&	7.7	$\pm$	2.6	&	32.8	$\pm$	2.7	\\
HD  48915	&	2018-08-27 18:57:03	&	2018AUG	&	HIPPI-2	&	11.9	&	E6	&	500SP	&	B	&	438.6	&	0.467	&	10.2	$\pm$	1.5	&	0.0	$\pm$	1.6	\\
HD 102647	&	2018-07-11 09:14:15	&	2018JUL	&	HIPPI-2	&	11.9	&	E4	&	500SP	&	B	&	440.7	&	0.682	&	1.6	$\pm$	2.6	&	26.1	$\pm$	2.6	\\
HD 102647	&	2018-07-16 08:57:38	&	2018JUL	&	HIPPI-2	&	11.9	&	E4	&	500SP	&	B	&	440.7	&	0.682	&	-11.4	$\pm$	2.6	&	10.7	$\pm$	2.7	\\
HD 102870	&	2018-07-12 08:56:01	&	2018JUL	&	HIPPI-2	&	11.9	&	E4	&	500SP	&	B	&	446.1	&	0.716	&	-18.1	$\pm$	5.6	&	15.1	$\pm$	5.4	\\
HD 140573	&	2018-07-17 12:39:21	&	2018JUL	&	HIPPI-2	&	11.9	&	E4	&	500SP	&	B	&	450.5	&	0.753	&	-9.9	$\pm$	4.3	&	18.4	$\pm$	4.3	\\
HD 140573	&	2018-08-17 09:41:43	&	2018AUG	&	HIPPI-2	&	11.9	&	E5	&	500SP	&	B	&	450.1	&	0.636	&	-6.3	$\pm$	4.6	&	4.7	$\pm$	5.4	\\
\multicolumn{2}{l}{\textit{Adopted TP}}		&		&		&		&		&		&		&	445.0	&		&	2.5	$\pm$	1.4	&	18.4	$\pm$	1.5	\\
\hline																											
HD   2151	&	2018-07-12 18:14:27	&	2018JUL	&	HIPPI-2	&	11.9	&	E4	&	\gfil	&	B	&	471.0	&	0.829	&	-23.3	$\pm$	2.9	&	8.5	$\pm$	3.1	\\
HD   2151	&	2018-09-02 10:19:09	&	2018AUG	&	HIPPI-2	&	11.9	&	E7	&	\gfil	&	B	&	472.1	&	0.627	&	0.5	$\pm$	4.6	&	-35.3	$\pm$	4.9	\\
HD   2151	&	2018-09-02 16:21:37	&	2018AUG	&	HIPPI-2	&	11.9	&	E7	&	\gfil	&	B	&	471.0	&	0.620	&	-0.1	$\pm$	5.2	&	17.9	$\pm$	4.7	\\
HD  10700	&	2018-07-15 19:19:11	&	2018JUL	&	HIPPI-2	&	11.9	&	E4	&	\gfil	&	B	&	473.2	&	0.838	&	-11.6	$\pm$	4.3	&	4.7	$\pm$	4.2	\\
HD  10700	&	2018-08-28 15:35:45	&	2018AUG	&	HIPPI-2	&	11.9	&	E7	&	\gfil	&	B	&	473.2	&	0.633	&	-14.9	$\pm$	6.0	&	5.6	$\pm$	5.2	\\
HD  10700	&	2018-09-02 17:44:04	&	2018AUG	&	HIPPI-2	&	11.9	&	E7	&	\gfil	&	B	&	473.2	&	0.633	&	-13.7	$\pm$	5.8	&	10.3	$\pm$	5.9	\\
HD  48915	&	2018-08-17 19:40:14	&	2018AUG	&	HIPPI-2	&	11.9	&	E5	&	\gfil	&	B	&	464.4	&	0.704	&	-6.1	$\pm$	1.5	&	-2.0	$\pm$	1.4	\\
HD  48915	&	2018-08-20 19:36:03	&	2018AUG	&	HIPPI-2	&	11.9	&	E5	&	\gfil	&	B	&	464.4	&	0.704	&	-5.7	$\pm$	1.4	&	-2.5	$\pm$	1.4	\\
HD 102647	&	2018-07-10 09:16:20	&	2018JUL	&	HIPPI-2	&	11.9	&	E4	&	\gfil	&	B	&	466.0	&	0.809	&	-21.6	$\pm$	2.1	&	16.5	$\pm$	1.9	\\
HD 102647	&	2018-07-11 08:36:01	&	2018JUL	&	HIPPI-2	&	11.9	&	E4	&	\gfil	&	B	&	465.5	&	0.807	&	-14.1	$\pm$	1.9	&	10.5	$\pm$	2.0	\\
HD 102870	&	2018-07-12 08:18:28	&	2018JUL	&	HIPPI-2	&	11.9	&	E4	&	\gfil	&	B	&	470.4	&	0.827	&	-21.0	$\pm$	4.2	&	6.2	$\pm$	4.1	\\
HD 140573	&	2018-07-18 11:26:49	&	2018JUL	&	HIPPI-2	&	11.9	&	E4	&	\gfil	&	B	&	475.3	&	0.847	&	-19.8	$\pm$	3.1	&	10.2	$\pm$	3.1	\\
HD 140573	&	2018-08-16 09:50:15	&	2018AUG	&	HIPPI-2	&	11.9	&	E5	&	\gfil	&	B	&	475.3	&	0.764	&	-17.1	$\pm$	3.7	&	4.5	$\pm$	3.6	\\
\multicolumn{2}{l}{\textit{Adopted TP}}		&		&		&		&		&		&		&	470.4	&		&	-13.0	$\pm$	1.1	&	4.2	$\pm$	1.0	\\
\hline																											
HD   2151	&	2018-07-12 18:34:42	&	2018JUL	&	HIPPI-2	&	11.9	&	E4	&	V	&	B	&	536.7	&	0.951	&	-29.3	$\pm$	4.9	&	8.1	$\pm$	4.9	\\
HD  10700	&	2018-07-15 18:59:46	&	2018JUL	&	HIPPI-2	&	11.9	&	E4	&	V	&	B	&	539.5	&	0.949	&	-29.9	$\pm$	6.4	&	-3.3	$\pm$	6.4	\\
HD  48915	&	2018-08-16 19:42:31	&	2018AUG	&	HIPPI-2	&	11.9	&	E5	&	V	&	B	&	533.1	&	0.952	&	-5.8	$\pm$	1.7	&	5.0	$\pm$	1.5	\\
HD  48915	&	2018-08-27 19:08:28	&	2018AUG	&	HIPPI-2	&	11.9	&	E6	&	V	&	B	&	533.1	&	0.934	&	-17.1	$\pm$	1.3	&	2.9	$\pm$	1.5	\\
HD 102870	&	2018-07-17 08:42:01	&	2018JUL	&	HIPPI-2	&	11.9	&	E4	&	V	&	B	&	536.6	&	0.951	&	-32.2	$\pm$	3.6	&	-4.2	$\pm$	3.4	\\
HD 140573	&	2018-07-18 12:03:46	&	2018JUL	&	HIPPI-2	&	11.9	&	E4	&	V	&	B	&	541.3	&	0.948	&	-21.2	$\pm$	4.9	&	13.0	$\pm$	4.6	\\
HD 140573	&	2018-07-22 13:07:49	&	2018JUL	&	HIPPI-2	&	11.9	&	E4	&	V	&	B	&	541.6	&	0.948	&	-9.9	$\pm$	4.9	&	-6.6	$\pm$	4.4	\\
HD 140573	&	2018-08-17 09:59:15	&	2018AUG	&	HIPPI-2	&	11.9	&	E5	&	V	&	B	&	541.2	&	0.954	&	-19.6	$\pm$	4.9	&	2.4	$\pm$	4.6	\\
\multicolumn{2}{l}{\textit{Adopted TP}}		&		&		&		&		&		&		&	537.9	&		&	-20.6	$\pm$	1.6	&	2.2	$\pm$	1.5	\\
\hline																											
HD   2151	&	2018-07-23 17:54:47	&	2018JUL	&	HIPPI-2	&	11.9	&	E4	&	\rfil	&	R	&	625.6	&	0.823	&	-14.3	$\pm$	3.0	&	4.0	$\pm$	2.9	\\
HD   2151	&	2018-08-23 16:11:43	&	2018AUG	&	HIPPI-2	&	11.9	&	E5	&	\rfil	&	R	&	625.6	&	0.894	&	-14.8	$\pm$	3.5	&	-6.9	$\pm$	3.7	\\
HD   2151	&	2018-08-24 16:37:59	&	2018AUG	&	HIPPI-2	&	11.9	&	E6	&	\rfil	&	R	&	625.6	&	0.920	&	-12.8	$\pm$	3.8	&	-4.5	$\pm$	3.5	\\
HD   2151	&	2018-08-26 16:44:18	&	2018AUG	&	HIPPI-2	&	11.9	&	E6	&	\rfil	&	R	&	625.6	&	0.920	&	-16.8	$\pm$	3.1	&	-6.0	$\pm$	3.4	\\
HD  10700	&	2018-08-23 16:47:46	&	2018AUG	&	HIPPI-2	&	11.9	&	E5	&	\rfil	&	R	&	626.9	&	0.892	&	-14.9	$\pm$	4.3	&	1.5	$\pm$	4.3	\\
HD  10700	&	2018-08-24 17:24:12	&	2018AUG	&	HIPPI-2	&	11.9	&	E6	&	\rfil	&	R	&	626.9	&	0.918	&	-3.7	$\pm$	4.2	&	-0.4	$\pm$	4.3	\\
HD 102647	&	2018-07-23 08:50:14	&	2018JUL	&	HIPPI-2	&	11.9	&	E4	&	\rfil	&	R	&	622.9	&	0.828	&	-10.8	$\pm$	3.1	&	4.9	$\pm$	3.3	\\
HD 102647	&	2018-07-24 08:39:56	&	2018JUL	&	HIPPI-2	&	11.9	&	E4	&	\rfil	&	R	&	622.9	&	0.828	&	-1.6	$\pm$	3.2	&	4.5	$\pm$	3.1	\\
HD 102870	&	2018-07-23 09:10:07	&	2018JUL	&	HIPPI-2	&	11.9	&	E4	&	\rfil	&	R	&	625.7	&	0.823	&	-17.3	$\pm$	5.3	&	13.4	$\pm$	5.3	\\
HD 102870	&	2018-07-24 09:18:08	&	2018JUL	&	HIPPI-2	&	11.9	&	E4	&	\rfil	&	R	&	625.7	&	0.823	&	-13.7	$\pm$	5.6	&	0.6	$\pm$	5.5	\\
HD 140573	&	2018-08-28 09:48:59	&	2018AUG	&	HIPPI-2	&	11.9	&	E7	&	\rfil	&	R	&	627.8	&	0.956	&	-16.8	$\pm$	2.8	&	-4.2	$\pm$	2.6	\\
\multicolumn{2}{l}{\textit{Adopted TP}}		&		&		&		&		&		&		&	625.6	&		&	-12.5	$\pm$	1.2	&	0.6	$\pm$	1.2	\\
\hline																											
HD   2151	&	2018-08-23 16:28:43	&	2018AUG	&	HIPPI-2	&	11.9	&	E5	&	650LP	&	R	&	725.2	&	0.729	&	-12.5	$\pm$	6.4	&	-3.7	$\pm$	6.3	\\
HD   2151	&	2018-08-24 16:57:40	&	2018AUG	&	HIPPI-2	&	11.9	&	E6	&	650LP	&	R	&	725.2	&	0.766	&	-9.5	$\pm$	4.5	&	2.5	$\pm$	4.4	\\
HD   2151	&	2018-08-26 17:04:47	&	2018AUG	&	HIPPI-2	&	11.9	&	E6	&	650LP	&	R	&	725.2	&	0.766	&	-16.9	$\pm$	4.3	&	-4.3	$\pm$	4.3	\\
HD  10700	&	2018-08-23 17:08:19	&	2018AUG	&	HIPPI-2	&	11.9	&	E5	&	650LP	&	R	&	728.6	&	0.724	&	-11.5	$\pm$	5.5	&	-8.9	$\pm$	5.8	\\
HD  10700	&	2018-08-24 17:45:22	&	2018AUG	&	HIPPI-2	&	11.9	&	E6	&	650LP	&	R	&	728.6	&	0.761	&	-3.1	$\pm$	5.5	&	4.2	$\pm$	6.0	\\
HD 102647	&	2018-07-23 08:32:24	&	2018JUL	&	HIPPI-2	&	11.9	&	E4	&	650LP	&	R	&	721.4	&	0.647	&	-3.3	$\pm$	5.7	&	-1.9	$\pm$	5.4	\\
HD 102647	&	2018-07-24 08:22:46	&	2018JUL	&	HIPPI-2	&	11.9	&	E4	&	650LP	&	R	&	721.4	&	0.647	&	-14.9	$\pm$	5.7	&	-1.3	$\pm$	5.7	\\
HD 102870	&	2018-07-23 09:30:33	&	2018JUL	&	HIPPI-2	&	11.9	&	E4	&	650LP	&	R	&	725.0	&	0.637	&	8.3	$\pm$	9.1	&	12.3	$\pm$	9.1	\\
HD 102870	&	2018-07-24 08:59:24	&	2018JUL	&	HIPPI-2	&	11.9	&	E4	&	650LP	&	R	&	725.0	&	0.637	&	11.9	$\pm$	8.7	&	37.4	$\pm$	8.7	\\
HD 140573	&	2018-08-28 09:28:41	&	2018AUG	&	HIPPI-2	&	11.9	&	E7	&	650LP	&	R	&	731.5	&	0.851	&	-19.5	$\pm$	3.1	&	2.0	$\pm$	2.9	\\
\multicolumn{2}{l}{\textit{Adopted TP}}		&		&		&		&		&		&		&	725.7	&		&	-7.1	$\pm$	1.9	&	3.8	$\pm$	1.9	\\
\hline																											
    \hline
\end{tabular}
\end{table*}

\clearpage

\section{Dust composition models}
\label{app:composition_seds}

\begin{figure}
\centering
\includegraphics[width=0.5\textwidth]{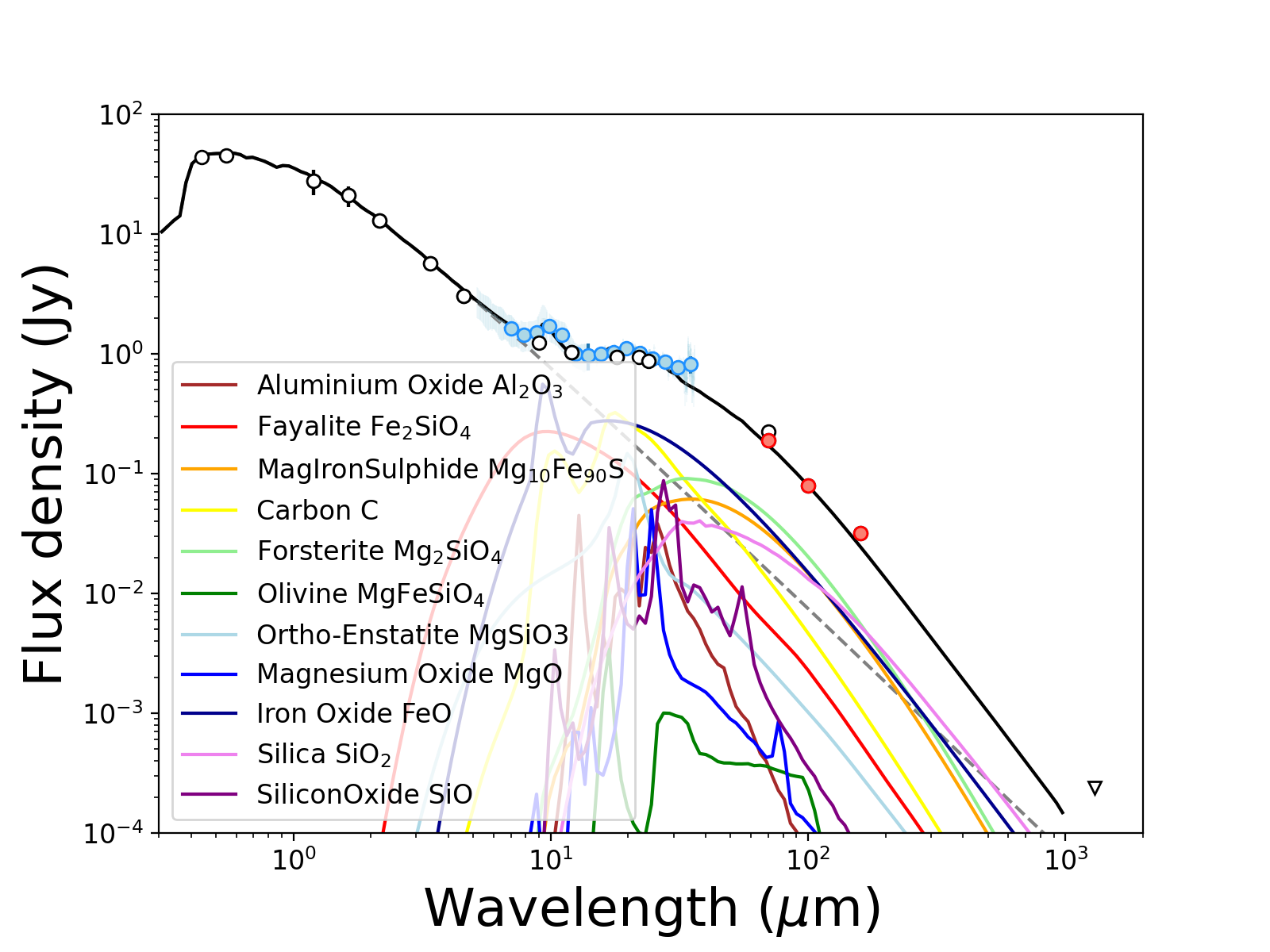}
\caption{Plot of flux density vs. wavelength for HD~172555. The stellar photosphere model (dashed black line), total model (solid black line), and observations (white, blue and red dots, with 1-$\sigma$ uncertainties) are presented alongside the weighted spectra for all 11 dust components fitted to the mid-infrared spectrum and far-infrared photometry. Brown: Al$_2$O$_3$, Red: C, Orange: Mg$_2$SiO$_4$, Yellow: MgFeSiO$_4$, Light Green: MgSiO$_3$, Green: MgO, Light Blue: FeO, Blue: SiO$_2$, Dark Blue: SiO, Violet: MgFeS, Purple: Fe$_2$SiO$_4$. \label{fig:sed_weights}}
\end{figure}

\begin{figure*}
\centering
\subfigure{\includegraphics[width=0.32\textwidth]{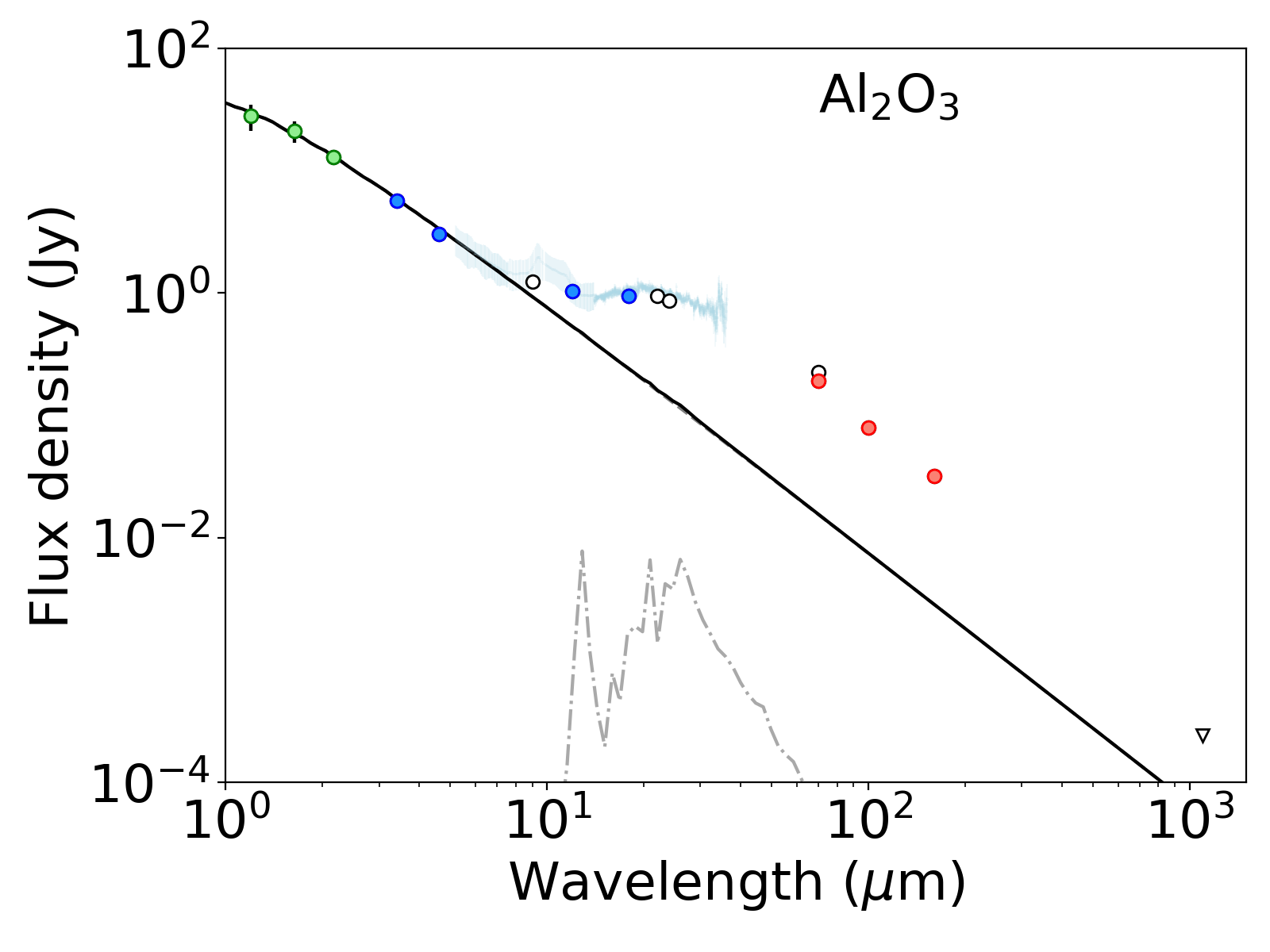}}
\subfigure{\includegraphics[width=0.32\textwidth]{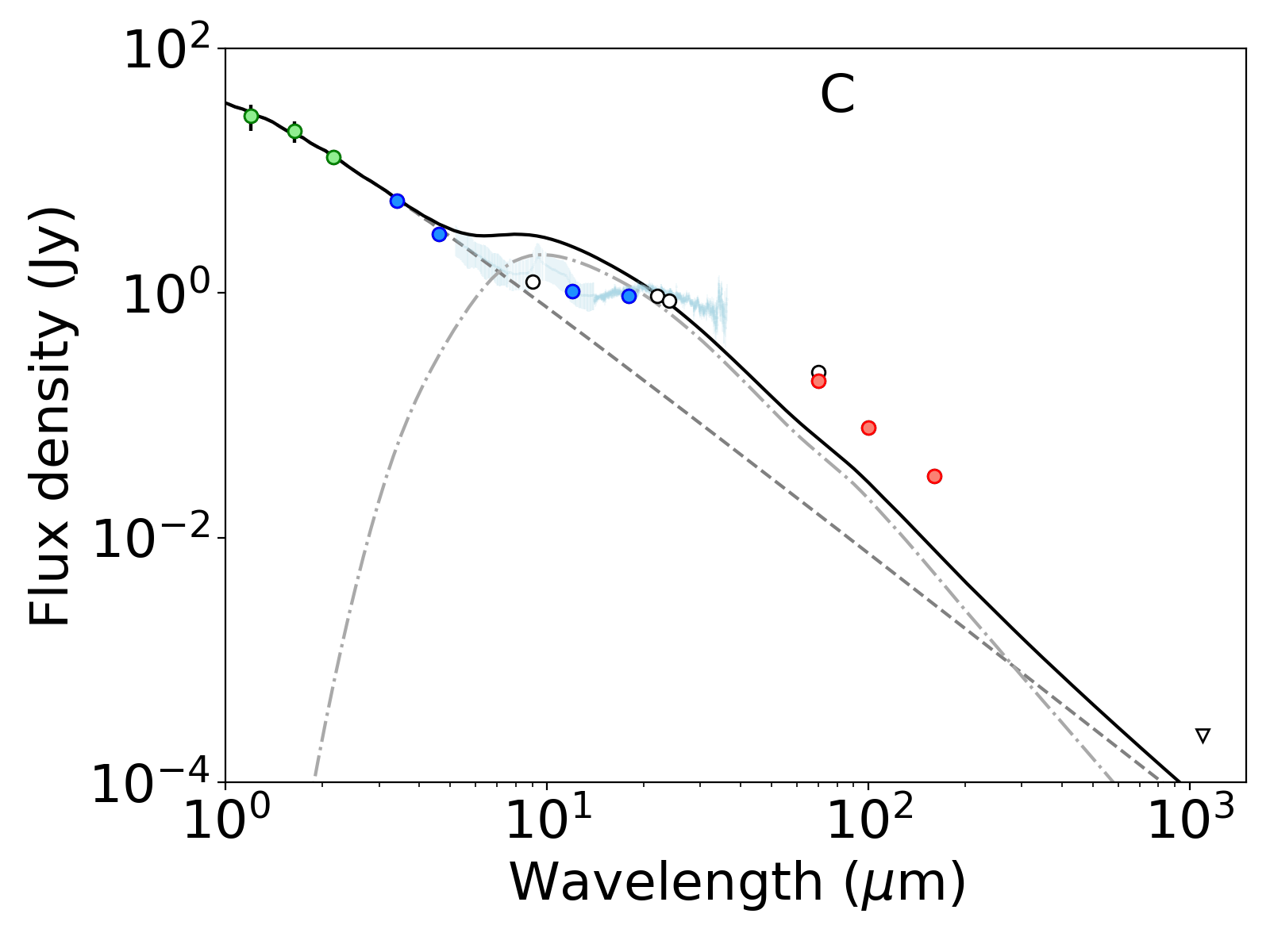}}
\subfigure{\includegraphics[width=0.32\textwidth]{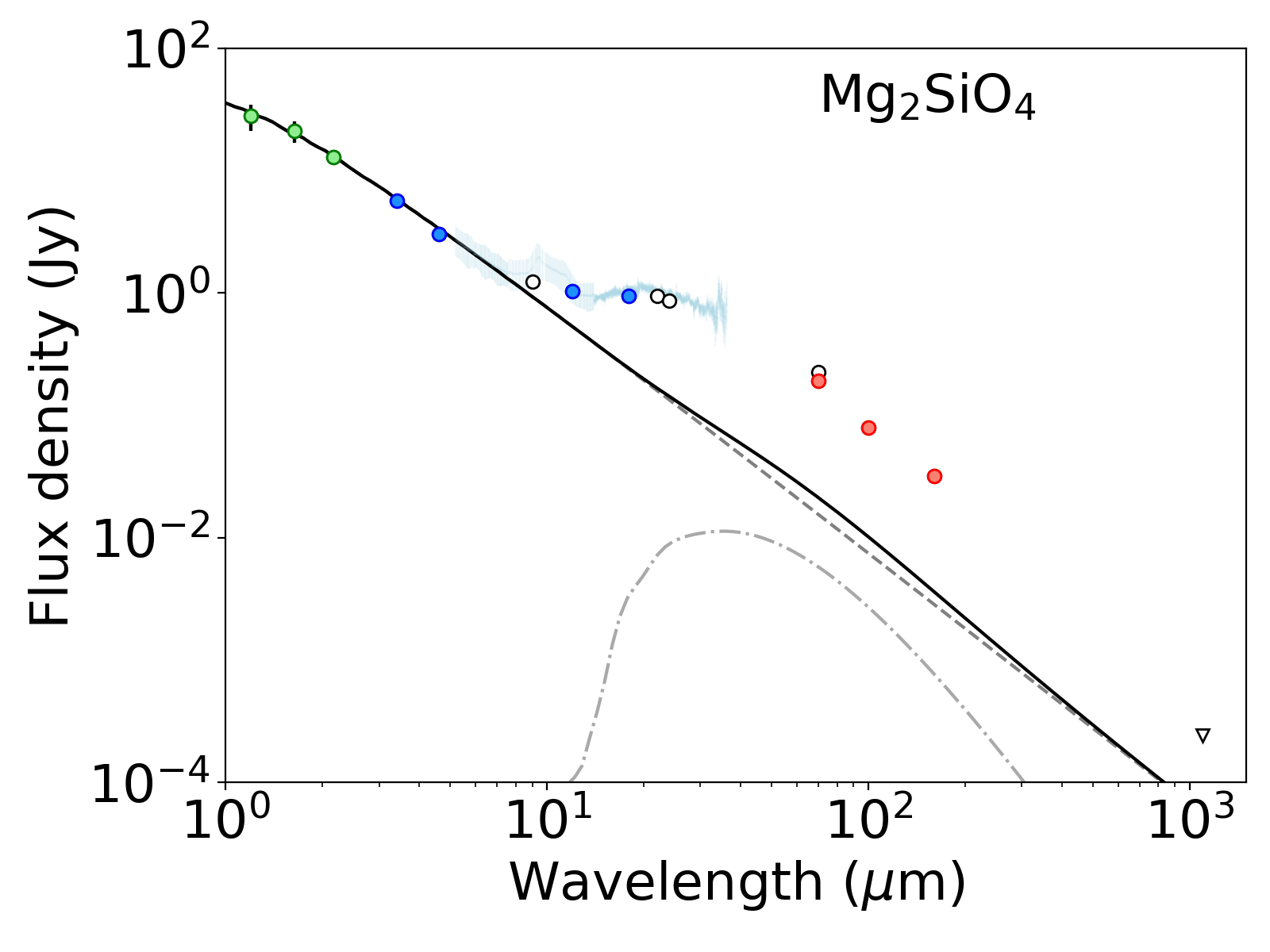}}\\
\subfigure{\includegraphics[width=0.32\textwidth]{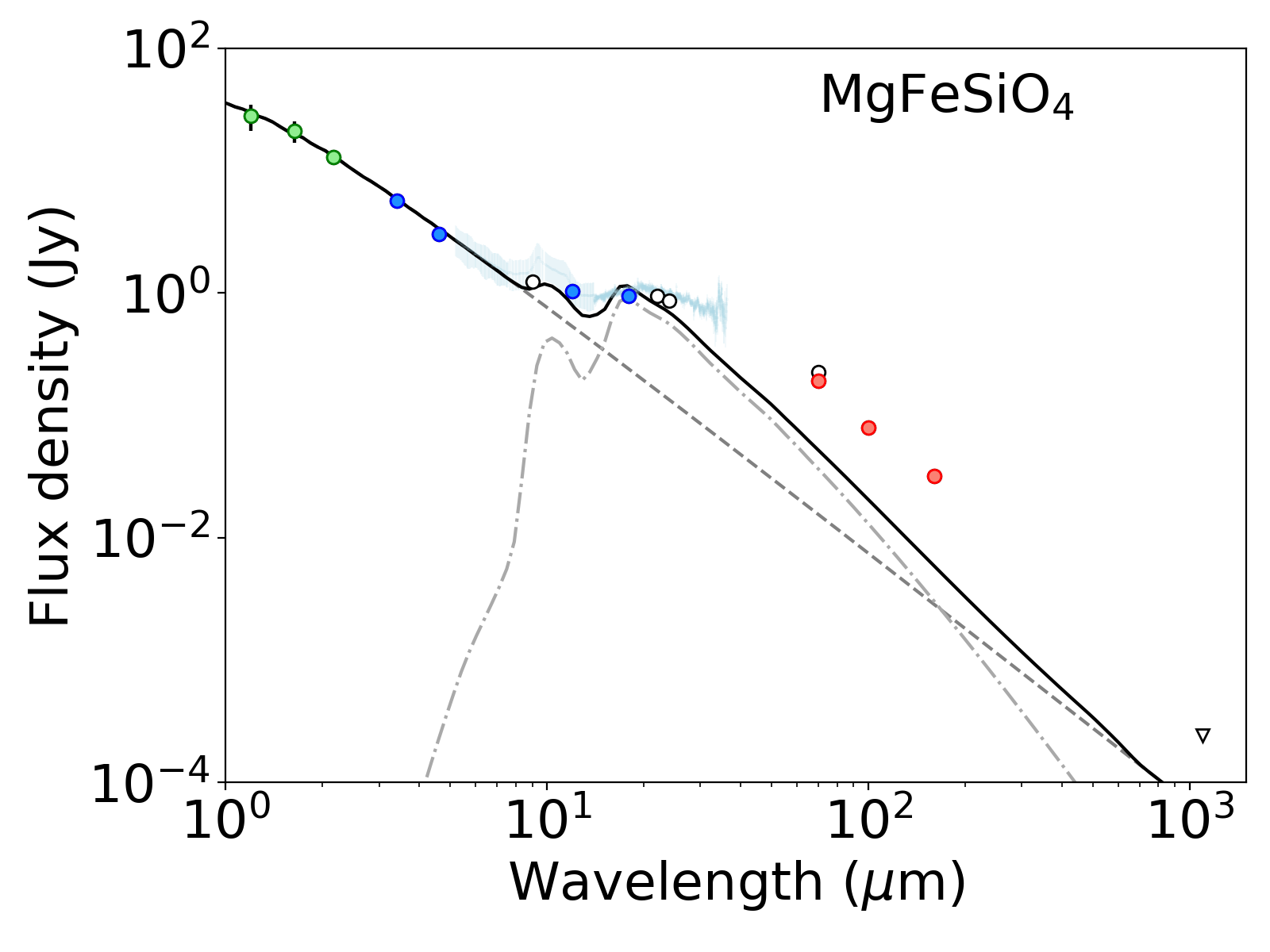}}
\subfigure{\includegraphics[width=0.32\textwidth]{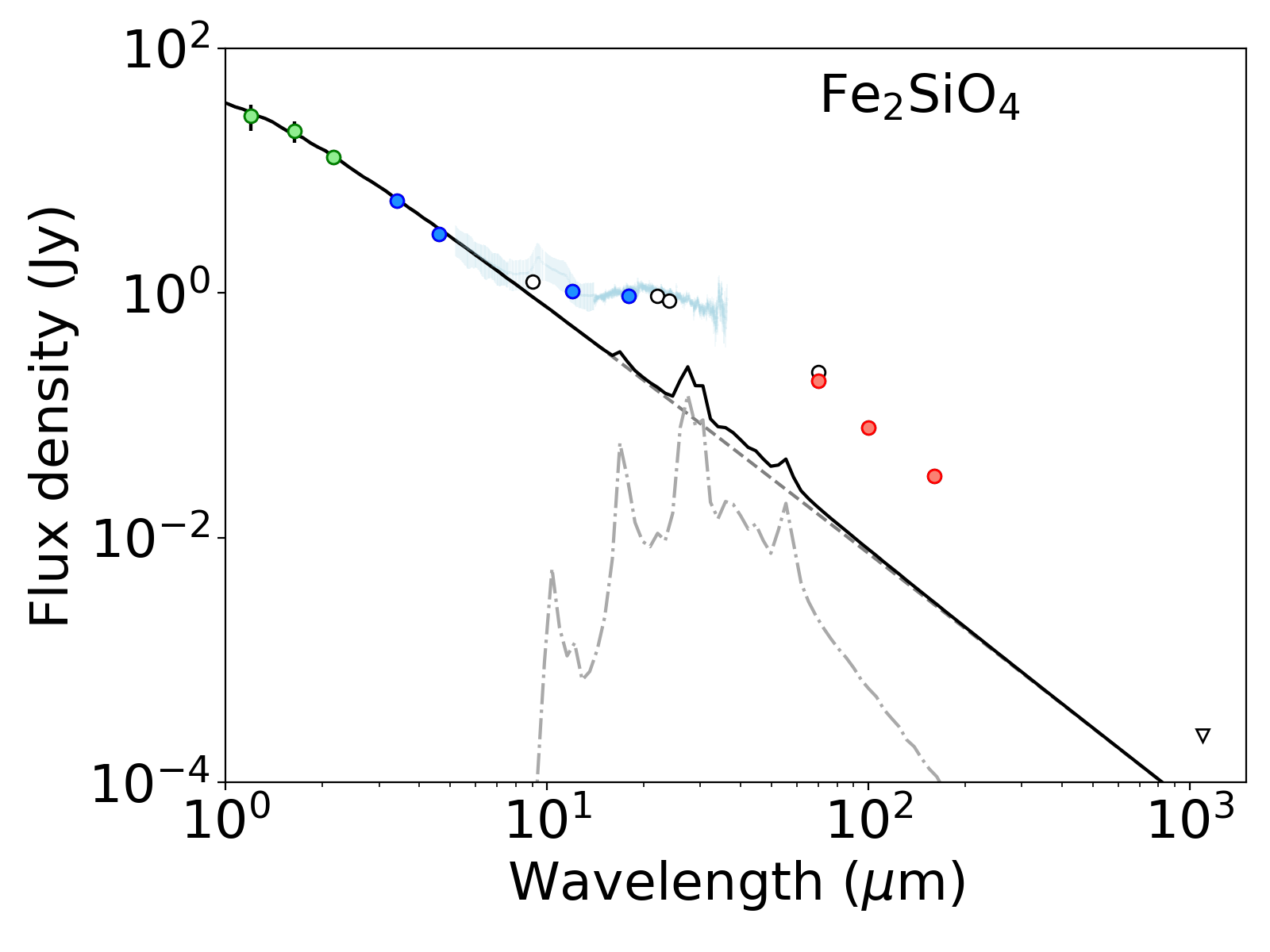}}
\subfigure{\includegraphics[width=0.32\textwidth]{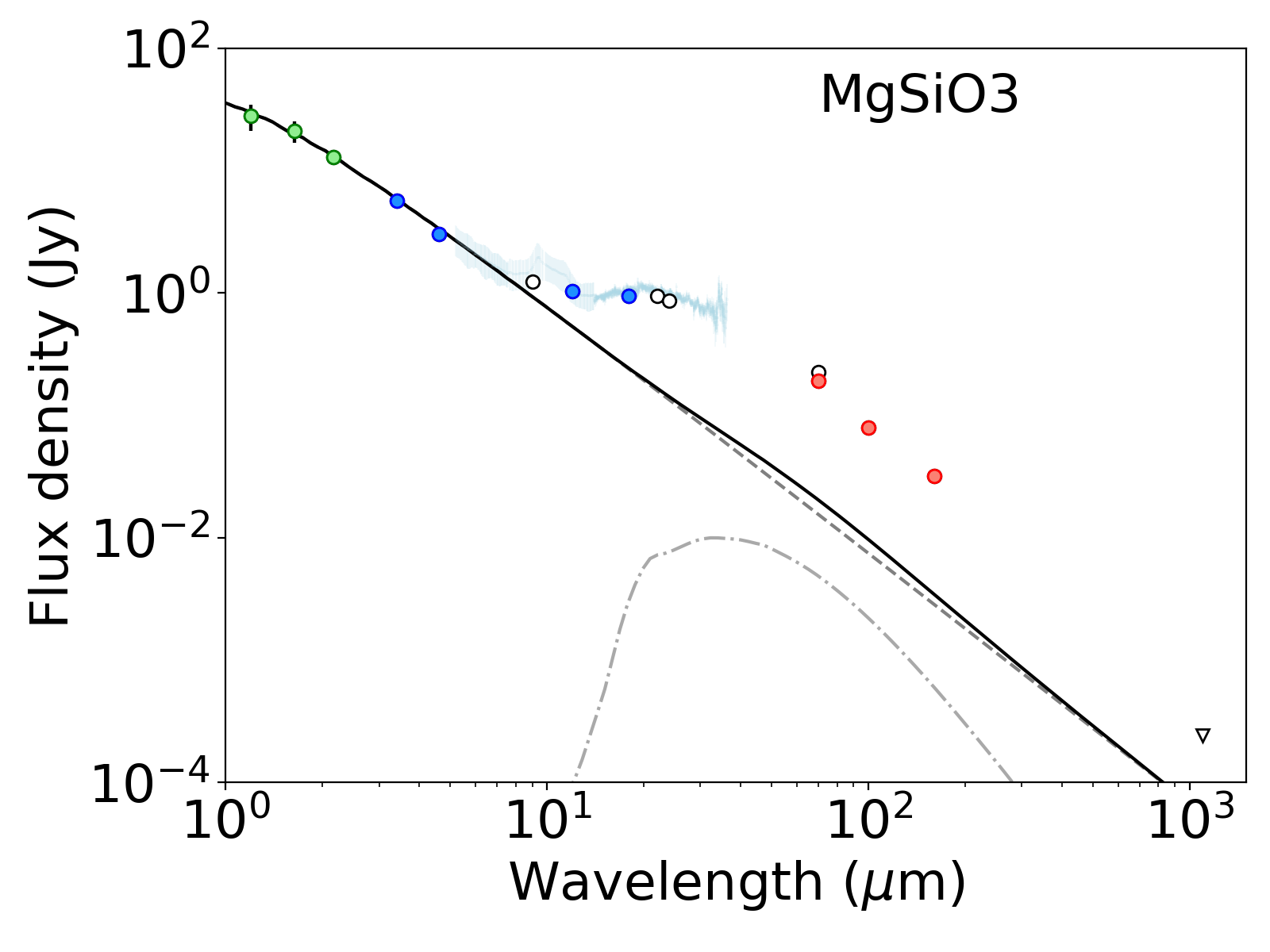}}\\
\subfigure{\includegraphics[width=0.32\textwidth]{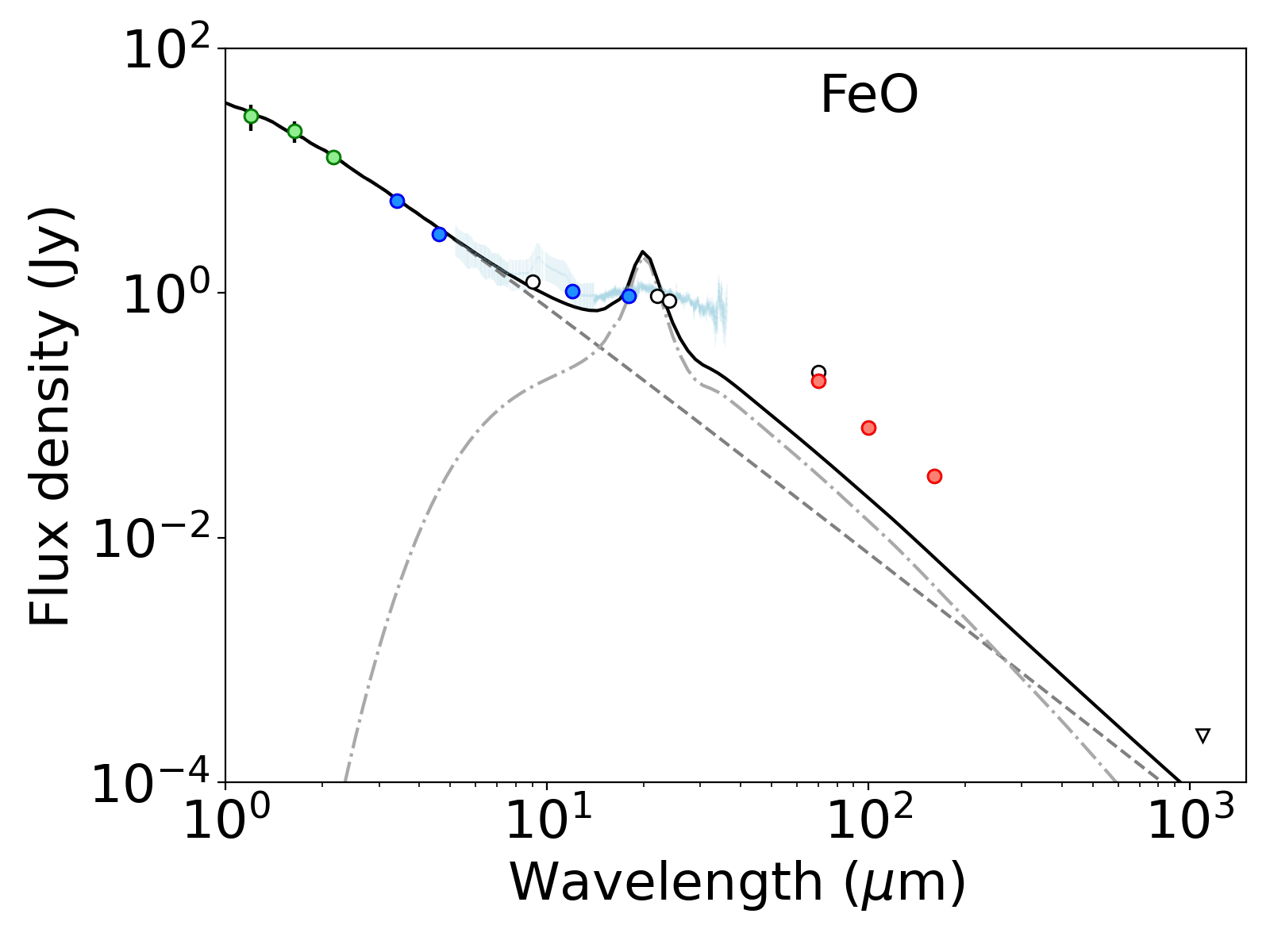}}
\subfigure{\includegraphics[width=0.32\textwidth]{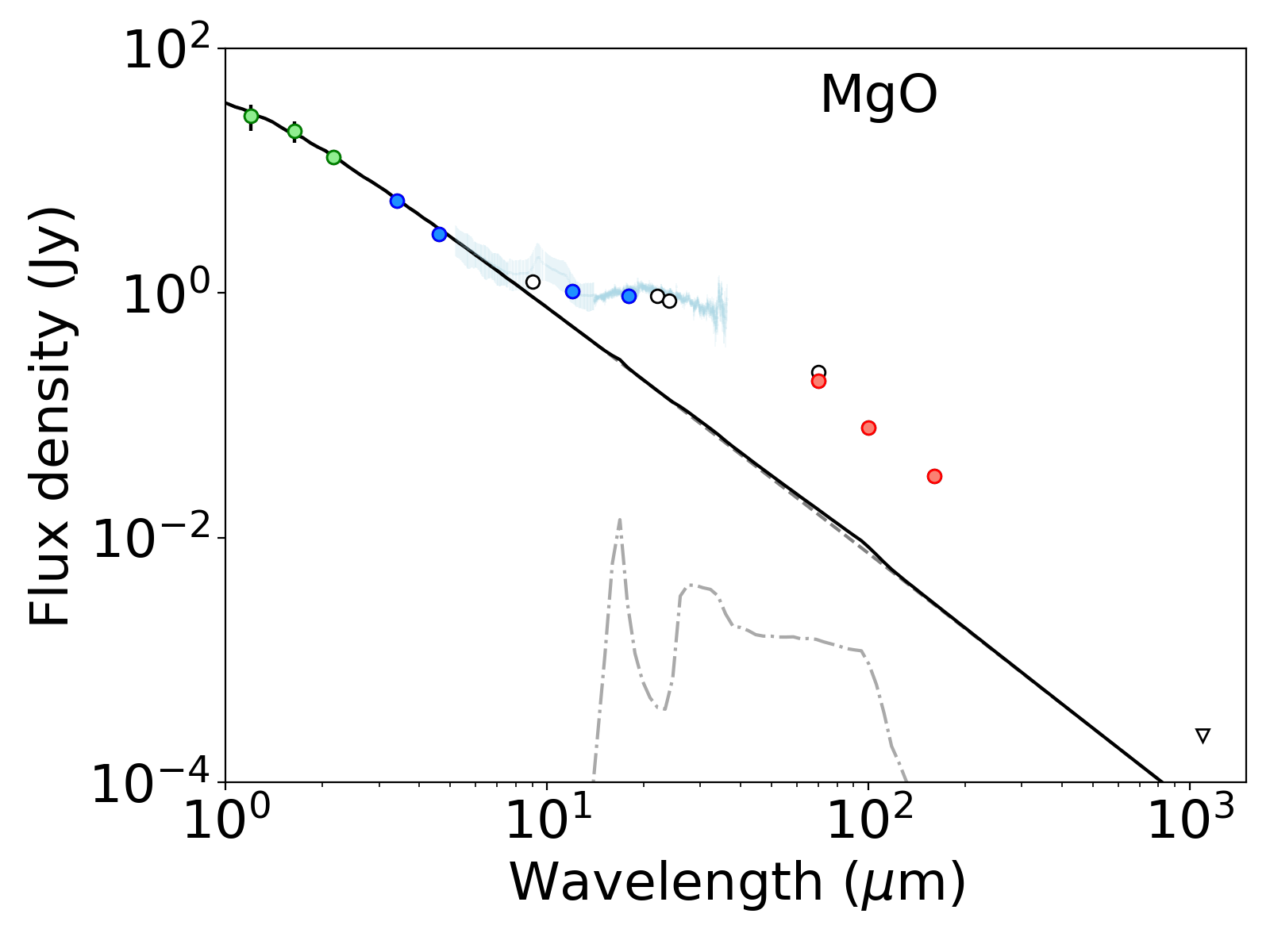}}
\subfigure{\includegraphics[width=0.32\textwidth]{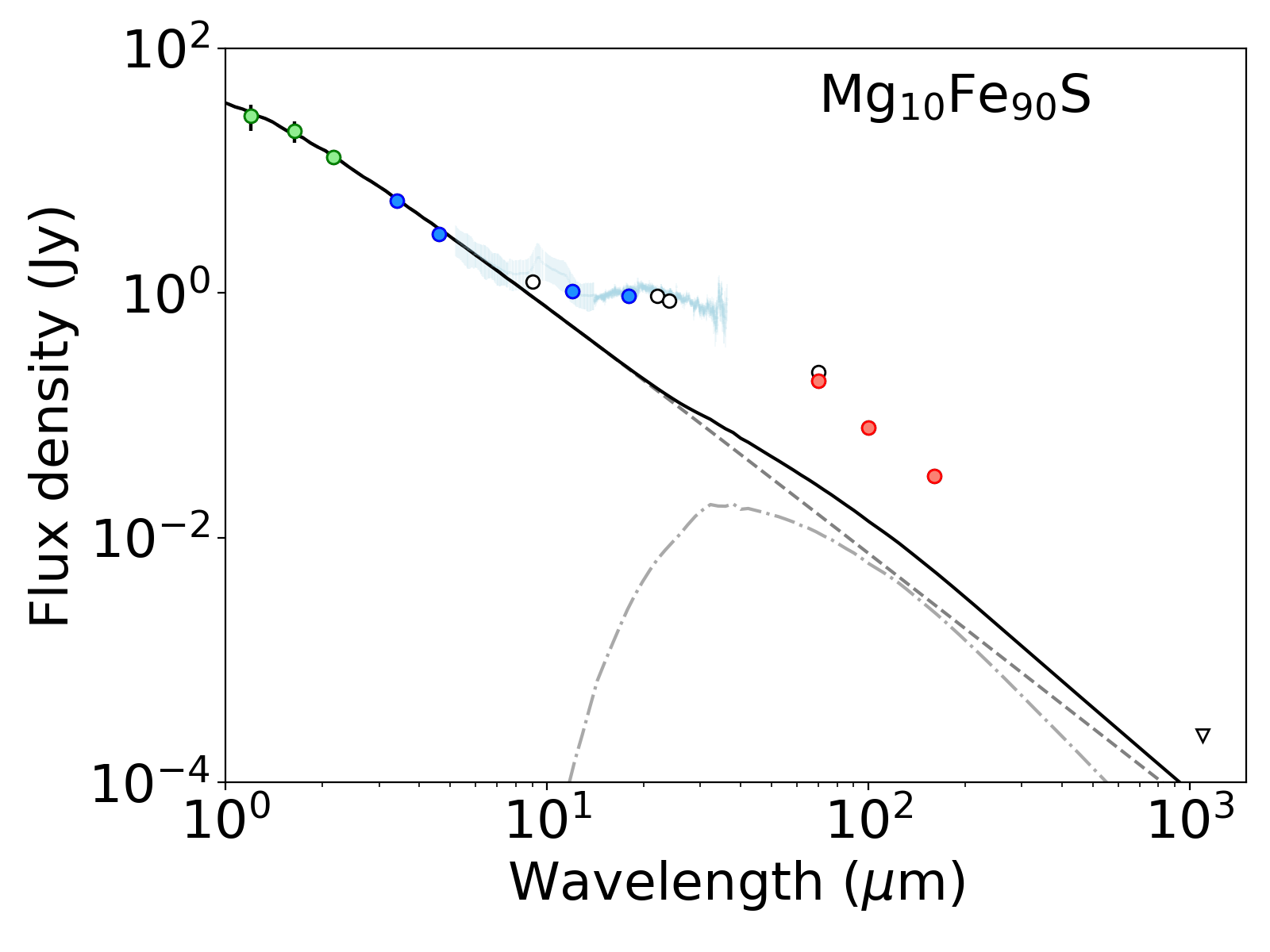}}\\
\subfigure{\includegraphics[width=0.32\textwidth]{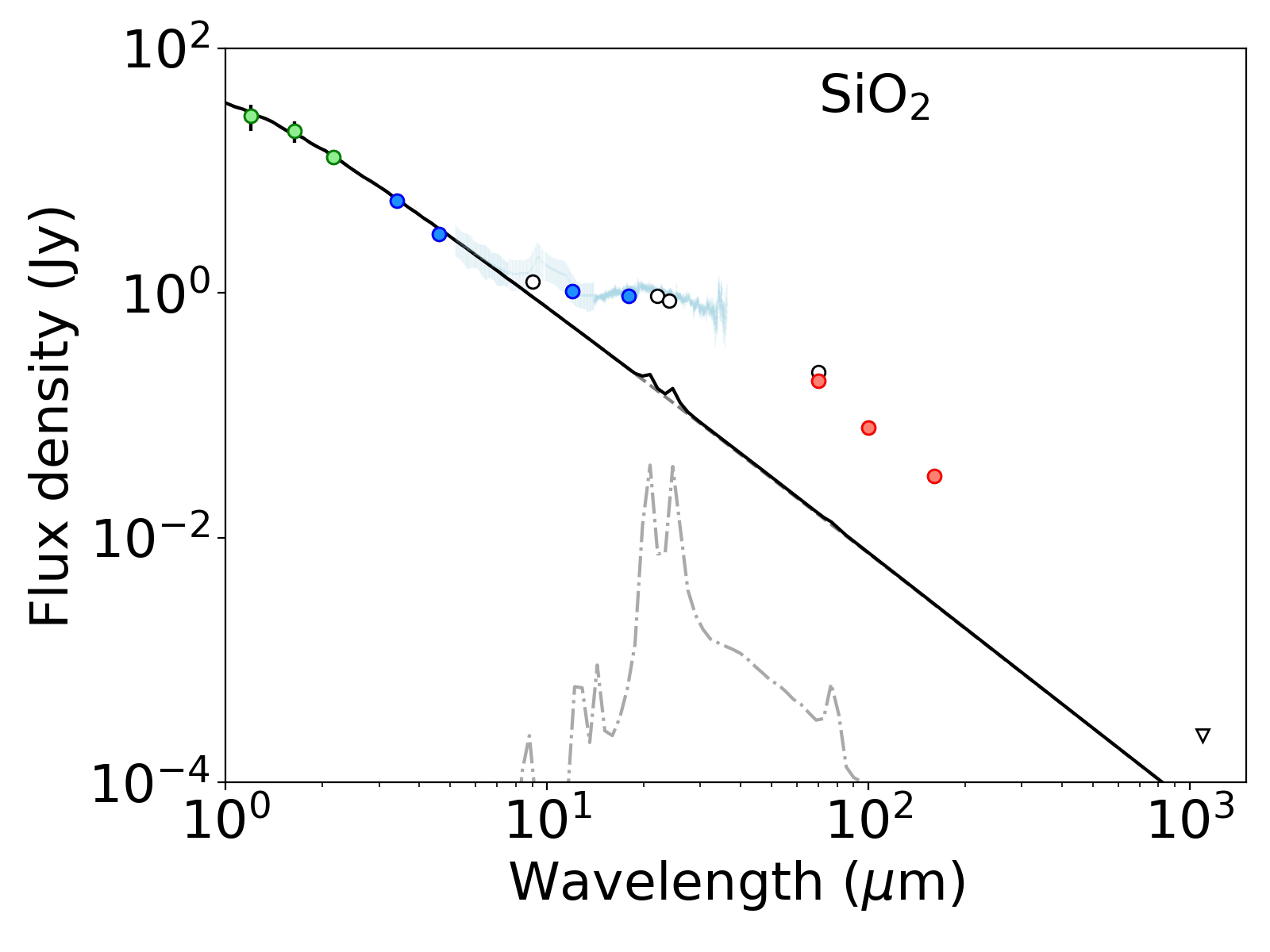}}
\subfigure{\includegraphics[width=0.32\textwidth]{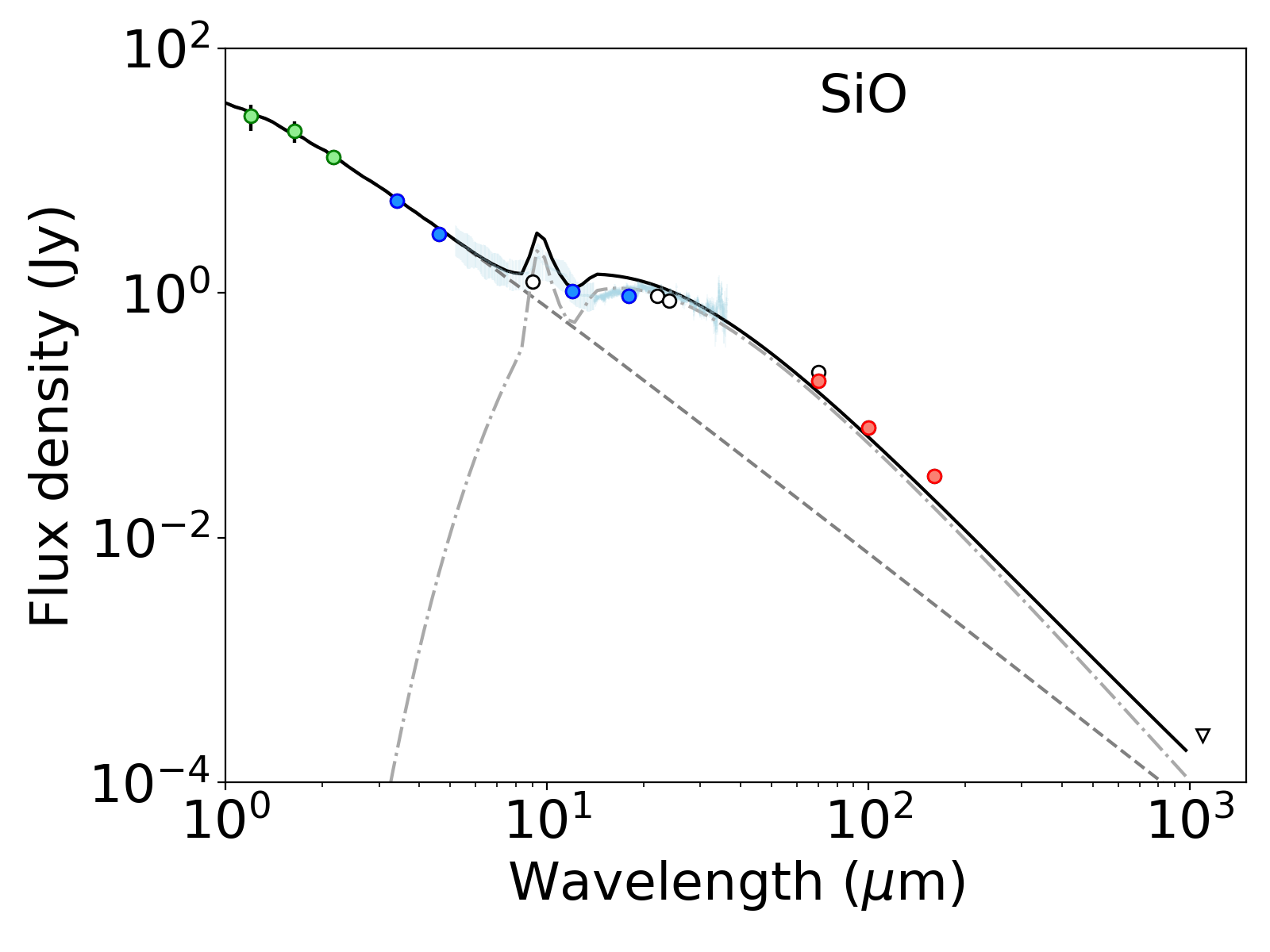}}\\
\caption{Spectral energy distributions for the eleven individual mineral species used to model HD~172555's infrared excess. From top left to bottom right, the species are: Aluminium Oxide, Carbon, Forsterite, Olivine, Fayalite, Ortho-Enstatite, Iron Oxide, Magnesium Oxide, Magnesium Iron Sulphide, Silicon Oxide, and Silicon Dioxide. The grain size for each species spanned 0.01 to 1000~$\mu$m, with a power law size distribution exponent of -3.95; the spatial distribution was matched to the scattered light image of \citet{2018Engler}. Each SED is scaled to the same dust mass of 3.45$\times10^{20}$~kg, or $5.78\times10^{-5}~M_{\odot}$. The choice of mass for the SED scaling was arbitrary. Observations are shown by circular data points (upper limit an inverted triangle), as per the main text. The grey dashed line show the stellar photosphere, grey dot-dash line shows the dust contribution, and the black solid line is the sum of these components. \label{fig:sed_components}}
\end{figure*}

We take optical constants ($n$,$k$) for representative materials from the Jena Database of Optical Constants. These optical constants are extrapolated to cover the wavelength range appropriate for this work. An SED showing the weighted contributions in comparison with each other is presented in Figure \ref{fig:sed_weights}, whilst the individual reference SEDs used to calculate weightings for each component's contribution to the observed SED are presented in Figure \ref{fig:sed_components}. 

\bsp	
\label{lastpage}
\end{document}